\documentclass[twoside,11pt]{article}

\usepackage[abbrvbib, preprint]{jmlr2e}
\usepackage[T1]{fontenc}
\usepackage[utf8]{inputenc}
\usepackage[commentmarkup=footnote]{changes}
\usepackage{booktabs,multirow}
\usepackage{array}
\usepackage{adjustbox}
\usepackage{etoolbox}
\usepackage{xcolor}
\usepackage{amsmath,amssymb,mathtools}
\usepackage{tikz, pgfplots}
\usepackage{fontawesome5}
\usepackage{xparse}
\usepackage{standalone}
\usepackage[autoload=true, theme=default-plain]{jlcode}
\def\jlbasicfont{\ttfamily\relsize{-1}\selectfont}
\pgfplotsset{compat=1.9}
\usetikzlibrary{calc,pgfplots.colormaps,backgrounds,intersections}
\usepgfplotslibrary{groupplots}
\usetikzlibrary{matrix}
\definecolor{TolBrightBlue}{HTML}{4477AA}
\definecolor{TolBrightCyan}{HTML}{66CCEE}
\definecolor{TolBrightGreen}{HTML}{228833}
\definecolor{TolBrightYellow}{HTML}{CCBB44}
\definecolor{TolBrightRed}{HTML}{EE6677}
\definecolor{TolBrightPurple}{HTML}{AA3377}
\definecolor{TolBrightGray}{HTML}{BBBBBB}

\definecolor{TolVibrantBlue}{HTML}{0077BB}
\definecolor{TolVibrantCyan}{HTML}{33BBEE}
\definecolor{TolVibrantTeal}{HTML}{009988}
\definecolor{TolVibrantOrange}{HTML}{EE7733}
\definecolor{TolVibrantRed}{HTML}{CC3311}
\definecolor{TolVibrantMagenta}{HTML}{EE3377}
\definecolor{TolVibrantGray}{HTML}{BBBBBB}

\definecolor{TolMutedBlue}{HTML}{332288}
\definecolor{TolMutedCyan}{HTML}{88CCEE}
\definecolor{TolMutedTeal}{HTML}{44AA99}
\definecolor{TolMutedGreen}{HTML}{117733}
\definecolor{TolMutedOlive}{HTML}{999933}
\definecolor{TolMutedSand}{HTML}{DDCC77}
\definecolor{TolMutedRose}{HTML}{CC6677}
\definecolor{TolMutedWine}{HTML}{882255}
\definecolor{TolMutedPurple}{HTML}{AA4499}
\definecolor{TolVibrantTeal}{HTML}{009988}
\definecolor{TolBrightGray}{HTML}{BBBBBB}

\tikzset{maxacc/.style={TolBrightGray, ycomb, mark size=0pt}}
\tikzset{maxtime/.style={TolBrightGray, ycomb, mark size=0pt}}
\tikzset{geomstats/.style={draw=TolMutedBlue,thick}}
\tikzset{autograd/.style={}}
\tikzset{numpy/.style={dashed}}
\tikzset{pytorch/.style={dotted}}
\tikzset{tensorflow/.style={dash dot}}
\tikzset{geoopt/.style={TolMutedCyan, thick}}
\tikzset{manifolds/.style={TolVibrantTeal, thick}}
\tikzset{manopt/.style={TolMutedGreen, thick}}
\tikzset{pymanopt/.style={TolMutedOlive, thick}}
\tikzset{riemopt/.style={TolMutedSand, thick}}
\tikzset{roptlib/.style={TolMutedRose, thick}}
\pgfplotsset{every axis title/.style={below right,at={(0,1)}}}

\pgfplotstableread[col sep=tab]{pic/benchmark-plots/distance-euclidean-by-dim.data}\ManRnDist
\pgfplotstableread[col sep=tab]{pic/benchmark-plots/exp-euclidean-by-dim.data}\ManRnExp
\pgfplotstableread[col sep=tab]{pic/benchmark-plots/log-euclidean-by-dim.data}\ManRnLog
\pgfplotstableread[col sep=tab]{pic/benchmark-plots/distance-hyperbolic-by-dim.data}\ManHnDist
\pgfplotstableread[col sep=tab]{pic/benchmark-plots/exp-hyperbolic-by-dim.data}\ManHnExp
\pgfplotstableread[col sep=tab]{pic/benchmark-plots/log-hyperbolic-by-dim.data}\ManHnLog
\pgfplotstableread[col sep=tab]{pic/benchmark-plots/distance-sphere-by-dim.data}\ManSnDist
\pgfplotstableread[col sep=tab]{pic/benchmark-plots/exp-sphere-by-dim.data}\ManSnExp
\pgfplotstableread[col sep=tab]{pic/benchmark-plots/log-sphere-by-dim.data}\ManSnLog
\pgfplotstableread[col sep=tab]{pic/benchmark-plots/distance-rotations-by-dim.data}\ManSODist
\pgfplotstableread[col sep=tab]{pic/benchmark-plots/exp-rotations-by-dim.data}\ManSOExp
\pgfplotstableread[col sep=tab]{pic/benchmark-plots/log-rotations-by-dim.data}\ManSOLog
\pgfplotstableread[col sep=tab]{pic/benchmark-plots/distance-spd-by-dim.data}\ManSPDDist
\pgfplotstableread[col sep=tab]{pic/benchmark-plots/exp-spd-by-dim.data}\ManSPDExp
\pgfplotstableread[col sep=tab]{pic/benchmark-plots/log-spd-by-dim.data}\ManSPDLog
\pgfplotstableread[col sep=tab]{pic/benchmark-plots/distance-power-by-dim.data}\ManPowDist
\pgfplotstableread[col sep=tab]{pic/benchmark-plots/exp-power-by-dim.data}\ManPowExp
\pgfplotstableread[col sep=tab]{pic/benchmark-plots/log-power-by-dim.data}\ManPowLog

\pgfplotstableread[col sep=tab]{pic/accuracy-plots/distance-euclidean-by-dim.data}\ManRnDistAcc
\pgfplotstableread[col sep=tab]{pic/accuracy-plots/exp-euclidean-by-dim.data}\ManRnExpAcc
\pgfplotstableread[col sep=tab]{pic/accuracy-plots/log-euclidean-by-dim.data}\ManRnLogAcc
\pgfplotstableread[col sep=tab]{pic/accuracy-plots/distance-hyperbolic-by-dim.data}\ManHnDistAcc
\pgfplotstableread[col sep=tab]{pic/accuracy-plots/exp-hyperbolic-by-dim.data}\ManHnExpAcc
\pgfplotstableread[col sep=tab]{pic/accuracy-plots/log-hyperbolic-by-dim.data}\ManHnLogAcc
\pgfplotstableread[col sep=tab]{pic/accuracy-plots/distance-sphere-by-dim.data}\ManSnDistAcc
\pgfplotstableread[col sep=tab]{pic/accuracy-plots/exp-sphere-by-dim.data}\ManSnExpAcc
\pgfplotstableread[col sep=tab]{pic/accuracy-plots/log-sphere-by-dim.data}\ManSnLogAcc
\pgfplotstableread[col sep=tab]{pic/accuracy-plots/distance-rotations-by-dim.data}\ManSODistAcc
\pgfplotstableread[col sep=tab]{pic/accuracy-plots/exp-rotations-by-dim.data}\ManSOExpAcc
\pgfplotstableread[col sep=tab]{pic/accuracy-plots/log-rotations-by-dim.data}\ManSOLogAcc
\pgfplotstableread[col sep=tab]{pic/accuracy-plots/distance-spd-by-dim.data}\ManSPDDistAcc
\pgfplotstableread[col sep=tab]{pic/accuracy-plots/exp-spd-by-dim.data}\ManSPDExpAcc
\pgfplotstableread[col sep=tab]{pic/accuracy-plots/log-spd-by-dim.data}\ManSPDLogAcc
\pgfplotstableread[col sep=tab]{pic/accuracy-plots/distance-power-by-dim.data}\ManPowDistAcc
\pgfplotstableread[col sep=tab]{pic/accuracy-plots/exp-power-by-dim.data}\ManPowExpAcc
\pgfplotstableread[col sep=tab]{pic/accuracy-plots/log-power-by-dim.data}\ManPowLogAcc

\graphicspath{{pic/}}
\usepackage{siunitx}
\robustify\bfseries

\newcommand{\R}{\mathbb{R}}
\newcommand{\Rn}[1]{\R^{#1}}
\newcommand{\Cn}[1]{\mathbb{C}^{#1}}
\newcommand{\Hn}[1]{\mathcal{H}^{#1}}
\newcommand{\tspace}[2]{T_{#1}#2}

\tikzset{%
    /pgfplots/colormap/viridis,
    fillBGCMap/.style={/utils/exec={\pgfplotscolormapdefinemappedcolor{#1}}, fill=mapped color!66}%
}
\newcommand{\tabNode}[3]{\tikz[anchor=base, baseline]\node[fillBGCMap=#1, rounded corners=.3em, inner sep=.25em] (T) {\color{#2}#3};}
\NewDocumentCommand{\fN}{O{}}{\tabNode{1000}{gray}{\small \faTimes$^{\mathrm{#1}}$}}
\NewDocumentCommand{\fR}{O{}}{\tabNode{850}{black}{\small $\Rn{#1}$}}
\NewDocumentCommand{\fC}{O{}}{\tabNode{700}{black}{\small $\Cn{#1}$}}
\NewDocumentCommand{\fH}{O{}}{\tabNode{550}{black}{\small $\mathbb H^{#1}$}}
\NewDocumentCommand{\fP}{}{\tabNode{750}{black}{\small \faCheck}}
\NewDocumentCommand{\fM}{}{\tabNode{1000}{gray}{\small \faTimes}}

\newcommand{\retr}{\operatorname{retr}}
\newcommand{\cM}{\mathcal M}

\newcommand{\Sn}[1]{\mathbb{S}^{#1}}
\newcommand{\SOn}[1]{\mathrm{SO}(#1)}
\newcommand{\symposdef}[1]{\mathcal{P}(#1)}

\newcommand{\ip}[2]{\langle #1, #2 \rangle}
\newcommand{\partrans}[2]{\operatorname{PT}_{#2 \gets #1}}
\newcommand{\tT}{\mathrm{T}}

\newcommand\ie{i.\,e.\ }

\newcolumntype{L}[1]{>{\raggedright\let\newline\\\arraybackslash\hspace{0pt}}m{#1}}
\newcolumntype{C}[1]{>{\centering\let\newline\\\arraybackslash\hspace{0pt}}m{#1}}
\newcolumntype{R}[1]{>{\raggedleft\let\newline\\\arraybackslash\hspace{0pt}}m{#1}}

\newcolumntype{T}[2]{%
    >{\adjustbox{angle=#1,lap=\width-(#2)}\bgroup}%
    l%
    <{\egroup}%
}
\newcommand*\tHead{\multicolumn{1}{T{20}{1em}}}

\jmlrheading{1}{2021}{1-48}{4/00}{10/00}{axen21a}{Seth D. Axen, Mateusz Baran, Ronny Bergmann and Krzysztof Rzecki}

\ShortHeadings{Manifolds.jl: Data Analysis on Manifolds}{S.\,D.~Axen, M.~Baran, R.~Bergmann and K.~Rzecki}
\firstpageno{1}

\begin{document}

\title{Manifolds.jl: An Extensible Julia Framework for Data Analysis on Manifolds}

\author{\name Seth D. Axen \email seth@sethaxen.com \\
       \addr Cluster of Excellence Machine Learning: New Perspectives for Science\\
       University of Tübingen\\
       Maria-von-Linden-Str. 6, 72076 Tübingen, Germany
       \AND
       \name Mateusz Baran \email mbaran@agh.edu.pl \\
       \addr AGH University of Science and Technology \\
       30 Mickiewicz Ave., 30-059 Krakow, Poland \\
       \\
       Cracow University of Technology \\
       Faculty of Materials Science and Physics\\
       Podchor\k{a}\.{z}ych 1, 30-084 Krakow, Poland
       \AND
       \name Ronny Bergmann \email ronny.bergmann@ntnu.no\\
       \addr Department of Mathematical Sciences\\
       Norwegian University of Science and Technology\\
       NO-7041 Trondheim, Norway
       \AND
       \name Krzysztof Rzecki \email krz@agh.edu.pl \\
       \addr AGH University of Science and Technology \\
       30 Mickiewicz Ave., 30-059 Krakow, Poland \\
       \\
       Cracow University of Technology \\
       Faculty of Materials Science and Physics\\
       Podchor\k{a}\.{z}ych 1, 30-084 Krakow, Poland
       }

\maketitle

\begin{abstract}
	We present the Julia package \lstinline!Manifolds.jl!, providing a fast and easy-to-use library of Riemannian manifolds and Lie groups.
    This package enables working with data defined on a Riemannian manifold, such as the circle, the sphere, symmetric positive definite matrices, or one of the models for hyperbolic spaces.
	We introduce a common interface, available in \lstinline!ManifoldsBase.jl!, with which new manifolds, applications, and algorithms can be implemented.
	We demonstrate the utility of \lstinline!Manifolds.jl! using Bézier splines, an optimization task on manifolds, and principal component analysis on nonlinear data.
    In a benchmark, \lstinline!Manifolds.jl! outperforms all comparable packages for low-dimensional manifolds in speed;
    over Python and Matlab packages, the improvement is often several orders of magnitude, while over C/C++ packages, the improvement is two-fold.
    For high-dimensional manifolds, it outperforms all packages except for Tensorflow-Riemopt, which is specifically tailored for high-dimensional manifolds.
\end{abstract}

\section{Introduction}
\label{sec:introduction}

In many applications, measured data appears in non-linear spaces.
In such spaces, common data analysis operations like the computation of a mean are not directly possible, since for example there is no addition operation on such data.
However, many of these spaces have the advantage of being Riemannian manifolds, that is, they are smooth manifolds equipped with a Riemannian metric, \ie with a locally varying inner product~\cite{DoCarmo:1992,AbsilMahonySepulchre:2008,boumal2022intromanifolds}.
Such manifolds appear in many applications.
For example, in Interferometric Synthetic Aperture Radar (InSAR)~\cite{BuergmannRosenFielding:2000:1}, the measured data are angles, and in diffusion tensor magnetic resonance imaging (DT-MRI)~\cite{FletcherJoshi:2007:1, dryden_non-euclidean_2009,ghosh_riemannian_2008}, diffusion at each voxel is described by a symmetric positive definite matrix.
The special orthogonal group of rotations is used in analysis of data from orientation sensors~\citep{marins_extended_2001}.
The sphere is used in various applications of directional statistics such as hydrology~\citep{chen_new_2013}.
The discipline of computational anatomy~\citep{grenander_computational_1998,cruzorive_stereological_1985} extensively uses geometric methods, including Large Deformation Diffeomorphic Metric Mapping~\citep{miller_computational_2004} with applications to medical data analysis~\citep{pennec_riemannian_2019,tulik_use_2019}.
In image analysis, Grassmann and Stiefel manifolds are used for pose estimation by~\citet{turaga_statistical_2011}.
Various manifolds with Fisher-Rao-like metrics are also used for shape analysis by, for example, ~\citet{srivastava_shape_2011},~\citet{baran_closest_2018}, and~\citet{Rzecki:2021}.

Other commonly encountered spaces include spheres or the Stiefel manifold of matrices with orthonormal columns.

Using tools of differential geometry, many operations can be generalized to manifolds by using only generic functions like the inner product, its induced norm, geodesics, the exponential, or the logarithmic map.
Closed-form expressions for the generic functions do not always exist, but efficient implementations to approximate them are still often possible.
Recent advances continue to resolve numerical challenges with efficiently implementing these operations.
Our library offers a wide range of such implementations, which makes performant data analysis for manifold-valued data more accessible.

One example of an operation that can be realized with these generic functions is the mean, which can be rephrased as the point in the space that (globally) has the minimum sum of squared distances from the given data points.
This definition can even be generalized to metric spaces where this minimizer is called the Fréchet mean.
Local minimizers may also exist on a Riemannian manifold; these are the
Riemannian centers of mass~\citep{GroveKarcher:1973:1, Karcher:1977:1}, sometimes also called Karcher means.

A main necessity for a wider adoption of geometric methods is a common, extensible, and efficient collection of operations on a large variety of manifolds.
One of the first toolboxes for optimization on manifolds was the Matlab package Manopt~\citep{manopt}.
The manifolds within this package are objects, where a fixed metric is assumed
but not explicitly mentioned.
Implementing a second metric for the same manifold requires a new Matlab object.
A manifold can have an exponential map. While most have at least one retraction implemented, there is no generic way to choose the exponential map or one of the retractions while calling a solver.
Several libraries followed, such as the C++ library ROPTLIB~\citep{huang_roptlib_2018},
which does provide a short user manual with introductory examples and an interface overview but lacks comprehensive documentation.
The seven major libraries for differential geometry in Python are Pymanopt~\citep{townsend_pymanopt:_2016}, Geomstats~\citep{miolane_geomstats_2020}, Geoopt~\citep{kochurov_geoopt_2020}, TheanoGeometry~\citep{kuhnel_computational_2017}, Jax Geometry \citep{kuhnel_computational_2017}, Tensorflow RiemOpt~\citep{Smirnov:2021}, and McTorch~\citep{meghwanshi_mctorch_2018}, each with their own implementations of a sometimes limited set of Riemannian manifolds.
Geomstats implements a metric interface, such that different metrics on one manifold can be easily distinguished.
Some libraries are more focused on specific tasks.
For example, MVIRT~\citep{mvirt} provides a variational model formulation for manifold-valued image processing.
This toolbox introduces implementations of a cyclic proximal point algorithm~\citep{Bacak:2014:1} applied to first- and second-order total variation~\citep{BacakBergmannSteidlWeinmann:2016:1}, as well as total generalized variation~\citep{BergmannFitschenPerschSteidl:2018:1}.
MVIRT also provides the parallel Douglas–Rachford algorithm~\citep{BergmannPerschSteidl:2016:1} for large-scale, nonsmooth optimization.

The above packages all follow the same basic approach: they generically implement their statistical or optimization algorithms for an arbitrary Riemannian manifold and provide a certain set of Riemannian manifolds that implement the necessary properties for these generic implementations to work.
The speed of an algorithm then depends on efficiency of both the abstract implementation of the algorithm as well as of employed operations on the specific manifold considered in an application.

Here we present the Julia package \lstinline!Manifolds.jl!, together with its accompanying interface package \lstinline!ManifoldsBase.jl!.
They provide an easily extensible interface to implement methods on a generic Riemannian manifold.
Moreover, they implement a comprehensive, well-documented library of efficient operations on common manifolds.

The remainder of this paper is organized as follows:
Section~\ref{section:manifolds} provides basic definitions of concepts from differential geometry and illustrates them using the unit sphere.
Next, Section~\ref{section:structure-of-library} describes the general design of \lstinline!Manifolds.jl!.
Section~\ref{sec:related-work} compares features of this library to what other libraries offer.
A few examples of usage of \lstinline!Manifolds.jl! are given in Section~\ref{sec:examples}.
Section~\ref{sec:benchmarks} presents performance benchmarks of \lstinline!Manifolds.jl! and other libraries on a set of manifolds for selected operations.
Finally, conclusions are contained in Section~\ref{sec:conclusions}.

\section{Manifolds}
\label{section:manifolds}

In this section we introduce the necessary terms to work on Riemannian manifolds.
We refer the reader to the textbooks \cite{DoCarmo:1992,Lee:2012:1,AbsilMahonySepulchre:2008,boumal2022intromanifolds}.
After introducing manifolds and related terms and definitions, the concepts are illustrated
for the unit sphere in $\Rn{3}$ and the manifold of symmetric positive definite matrices.

\subsection{General definitions}
\label{section:general-definitions}

We denote by $\cM$ a $d$-dimensional manifold.
A manifold $\cM$ is defined using an atlas $\mathcal A$ consisting of charts $\varphi \colon U \to \Rn{d}$ that introduce bijections from subsets $U \subset \cM$ of $\cM$ to open subsets of $\Rn{d}$ \cite{AbsilMahonySepulchre:2008}.
One way to work on manifolds is to work in charts.
Often, however, it is more computationally convenient or efficient to represent points on a manifold using coordinates of an embedding, instead of using a chart and working with \emph{intrinsic} methods.
Intrinsic here refers to the idea that the methods and tools used are “well defined regardless of the embedding space”\cite[p.~3]{boumal2022intromanifolds}.

At every point $p\in\cM$ we denote the tangent space by $\tspace{p}{\cM}$ and the tangent vectors by $X_p,Y_p\in\tspace{p}{\cM}$, where we leave out the index when the point $p$ is clear from context.
We denote the Riemannian metric by $\ip{\cdot}{\cdot}_p \colon \tspace{p}{\cM}\times\tspace{p}{\cM} \to \R$.
The metric also introduces a norm for tangent vectors $X\in \tspace{p}{\cM}$, which we denote by $\lVert X\rVert_p$.
Note that a smooth manifold can be equipped with different metrics.

For $p,q\in\cM$ and $X\in \tspace{p}{\cM}$ we denote by $\gamma_{p,X}\colon I \to \cM$ the geodesic starting in $\gamma_{p,X}(0)=p$ with $\dot\gamma_{p,X}(0) = X$,
where $I$ is the maximal interval over which the geodesic can be defined~\cite[Cor.~4.28]{Lee:2018:1} and $\mathcal O_p = \{ X \in \tspace{p}{\cM} : \gamma_{p,X} \text{ is defined on an interval containing } [0,1]\}$.  
A shortest geodesic denoted by $\gamma_{p;q}(t)$ starts at $p$ at time $t=0$, reaches $\gamma_{p;q}(1)=q$ at $t=1$, and is length-minimizing.
Note that the shortest geodesic might not be unique.

We further denote by $\exp_p\colon \tspace{p}{\cM} \to \cM$ the exponential map.
We define the injectivity radius $\operatorname{inj}(p)$ as the supremum over all radii $r>0$
such that $\exp_p$ is a diffeomorphism on the open ball $B(p,r) = \{ X \in \tspace{p}{\cM}: \lVert X \rVert_p < r\}$.
We denote by $U_p = B(p, \operatorname{inj}(p))$ and introduce $\mathcal U_p = \exp_p(U) \subset\cM$.
By definition $\exp_p\big|_{U_p}\colon U_p \to \mathcal U_p$ is a diffeomorphism
and hence the inverse of $\exp_p\big|_{U_p}$ exists, called the logarithmic map.
Formally the logarithmic map is defined as~\cite[Def.~10.20]{boumal2022intromanifolds}
\begin{equation*}
    \log_pq = \operatorname*{arg\,min}_{X \in \mathcal O_p} \lVert X \rVert\quad \text{ subject to} \exp_pX=q
\end{equation*}

In some cases, the exponential and logarithmic maps can be derived in closed form, but otherwise they are expensive to evaluate.
Therefore, one often uses a retraction $\retr_p \colon \tspace{p}{\cM} \to \cM$ that approximates the exponential map to (at least) first order\cite[Def.~3.47]{boumal2022intromanifolds}.
When the retraction can be inverted, \ie from $p$ and $q$ we obtain $X$, this inverse retraction $\retr_p^{-1} \colon q=\retr_p X \mapsto X$ can be used to approximate the logarithmic map.
While the exponential map, and hence also the logarithmic map, depend on the metric, a retraction is defined by properties of the retraction along a curve $c(t) = \retr_p(tX)$ for $X\in\tspace{p}{\cM}$ and its derivative $c'(t)$ and hence is independent of the metric.

A tangent space can be seen as "the directions to walk in" from a point $p$.
Let $q$ be a second point on $\cM$ such that the shortest geodesic is unique.
For a tangent vector $X\in \tspace{p}{\cM}$ at $p$ we denote the parallel transport to $q$ along the shortest geodesic (assuming it is unique) by $\partrans{p}{q} X$.
Again, since this map might not be given in closed form or might be expensive to compute, it is often approximated with methods called vector transports.

Computationally, points $p\in\cM$ and tangent vectors $X\in \tspace{p}{\cM}$ are usually represented by arrays, \ie vectors or matrices.
This is realized usually by embedding a manifold into some larger space.
A smooth function $i\colon\cM \to \Rn{m}$ whose differential $Di(p)$ has rank $d$ for every $p\in\cM$
is called an immersion.
The space $\Rn{m}$ is then called the embedding space of $\cM$, and $\cM$ is an embedded manifold.

A manifold $\cM$ is called a Lie group when we have a group operation $\circ\colon \cM\times\cM\to\cM$
that is smooth and whose inverse map $p \mapsto p^{-1}$ is also smooth.

\subsection{Example: the two-dimensional unit sphere}
\label{section:example-sphere}

The unit sphere in the three-dimensional Euclidean space, denoted $\Sn{2}$, is an intuitive example of a Riemannian manifold.
Points $p\in \Sn{2}$ on the unit sphere have unit norm $\lVert p \rVert_2 = 1$.
Adding (or subtracting) two of these points yields a point with a non-unit norm, which therefore does not belong to $\Sn{2}$.
However, there is an intuitive way of reinterpreting addition and subtraction that makes an analogous operation on the sphere possible.

\begin{figure}
    \centering
    \begin{tikzpicture}
        \node[anchor=south west,inner sep=0] (image) at (0,0) {\includegraphics[width=0.5\textwidth]{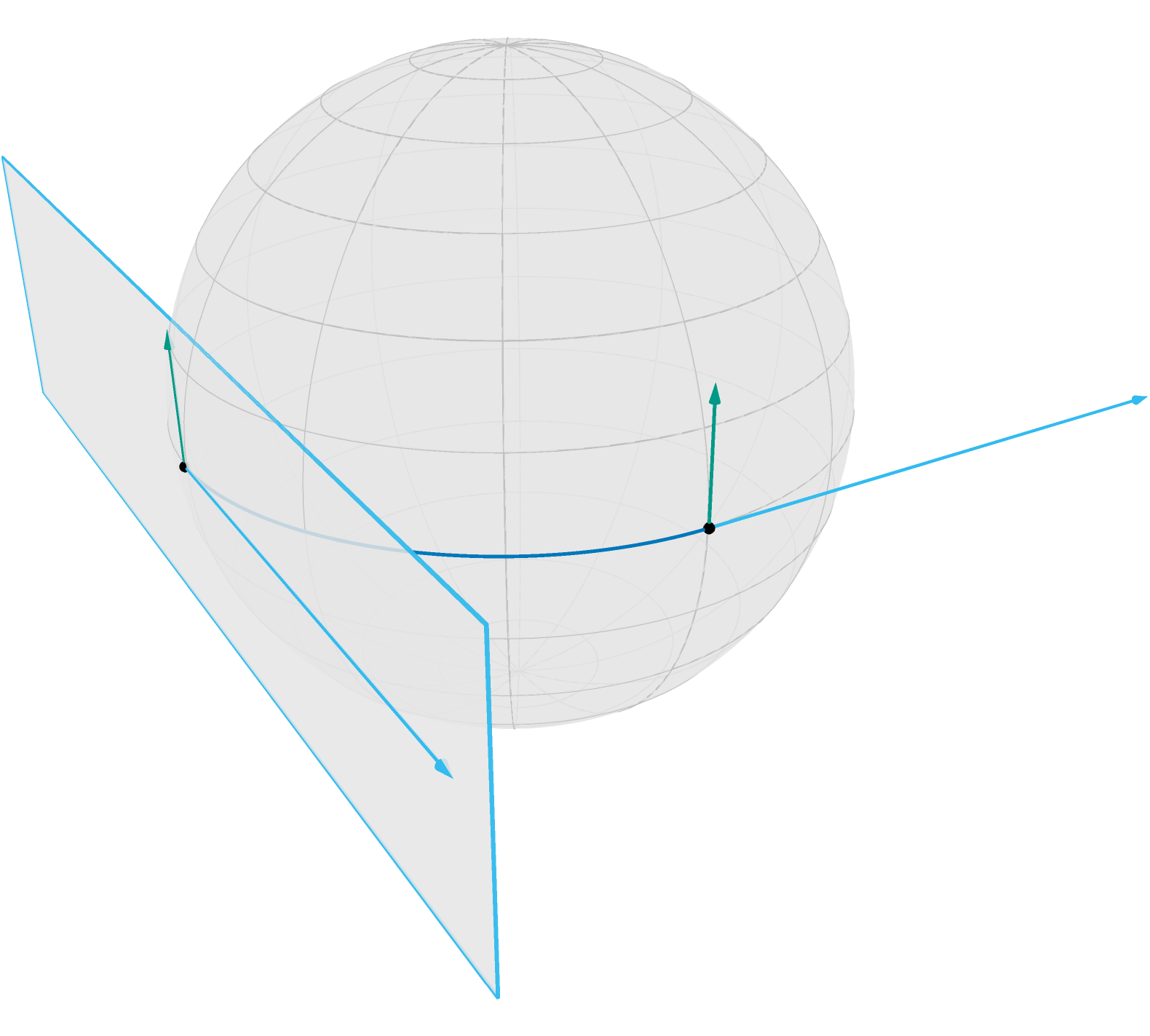}};
        \begin{scope}[x={(image.south east)},y={(image.north west)}]
            \node at (0.12, 0.56)  (p) {$p$};
            \node at (0.62, 0.45)  (q) {$q$};
            \node[text=TolVibrantTeal] at (0.14, 0.75)  (Y) {$Y$};
            \node[text=TolVibrantTeal] at (0.62, 0.66)  (Yq) {$\partrans{p}{q} Y$};
            \node[text=TolVibrantCyan] at (0.35, 0.35)  (X) {$X$};
            \node[text=TolVibrantCyan] at (0.85, 0.62)  (Xq) {$\partrans{p}{q} X$};
            \node[text=TolVibrantBlue] at (0.45, 0.5)  (g) {$\gamma(p,q)$};
            \node[text=TolVibrantGray] at (0.75, 0.35)  (M) {$\Sn{2}$};
            \node[text=TolVibrantCyan] at (0.5, 0.1)  (T) {$\tspace{p}{\Sn{2}}$};
        \end{scope}
    \end{tikzpicture}
    \caption{Unit sphere in $\Rn{3}$ with two points $p$, $q$.
    Tangent space at $p$ is labelled $\tspace{p}{\Sn{2}}$, geodesic between $p$ and $q$ is the blue arc $\gamma(p, q)$, and the vector $X$ is the tangent vector from $\tspace{p}{\Sn{2}}$ that points towards $q$.
    $Y$ is another vector from $\tspace{p}{\Sn{2}}$ orthogonal to $X$.
    All tangent vectors are blue.
    Parallel transports of $X$ and $Y$ from $p$ to $q$ are shown as vector $\partrans{p}{q} X$ and $\partrans{p}{q} Y$, respectively.}
    \label{fig:sphere}
\end{figure}

We can interpret the difference $y-x$ between two points $x,y \in \Rn{3}$ as a vector starting from $x$ and pointing to $y$.
Multiplying such a vector by a scalar and adding it to $x$ defines a set of points on a straight line passing through $x$ and $y$, i.\,e.\ $t \mapsto x+t(y-x)$.
This line segment joining $x$ and $y$ is the shortest connecting curve on $\Rn{3}$ called a shortest geodesic $\gamma_{x;y}(t)$ on $\Rn{3}$.
On $\Sn{2}$, the shortest geodesic is no longer a line segment but is instead a shortest great arc passing through points $p,q \in \Sn{2}$.
For any two non-opposing points, the intersection of their common plane through the origin with the sphere yields two geodesics, one shorter than the other.
We call this shorter great arc the shortest geodesic.
For two opposing points $p,-p \in \Sn{2}$, all great arcs that join the two points are half-circles; they are therefore joined by infinitely many shortest geodesics $\gamma_{p;-p}$.
We can measure the distance $d_{\Sn{2}}(p, q)$ between the points $p$ and $q$ as the length of a shortest geodesic, depicted as the blue curve in Figure~\ref{fig:sphere}.

To conceptualize tangent vectors, consider the set of vectors orthogonal to a point $p\in \Sn{2}$, that is
\begin{equation*}
    \tspace{p}{\Sn{2}} \coloneqq
    \bigl\{ X \in \Rn{3} \big| \langle X, p\rangle = 0 \bigr\}.
\end{equation*}
We obtain an actual plane tangent to the sphere if we consider all points $p+X$, $X\in \tspace{p}{\Sn{2}}$, which is tangent to the sphere and “touches” the sphere at $p$ (at $X=0$).
For every $X\in\tspace{p}\Sn{2}$ there exists a great arc $c(t)$, which is a geodesic (acceleration free curve when “measuring” in the tangent spaces it passes through) that runs through $c(0)=p$ with derivative $\dot c(0) = X$.
When parametrising the great arc with constant speed $\lVert \dot c(t) \rVert_2 = \lVert X \rVert_2$ we reach some point $c(1) = q \in \Sn{2}$ after time $t=1$. This point is $q=\exp_p X$.
As long as the norm of $X$ is less than $\pi$, the inverse function mapping $q$ to $X=\log_pq$ is unique as well.
With a start point $p$ and an initial velocity $X$, the great arc $c(t)$ can also be determined if both end points $c(0)=p$ and $c(1)=q$, $p \neq -q$ are given. The plane containing the origin, $p$, and $q$ is uniquely determined, and its intersection with the sphere $\Sn{2}$ forms a circle. The shorter arc segment connecting $p$ and $q$ on this circle is the image of the curve $c(t)$. Finally, there exists a unique constant speed parametrisation $\lVert \dot c(t)\rVert_2 = \mathrm{const}$ that reaches $q=c(1)$ after time $1$.
Then $X = \dot c(0) = \log_pq$.
If $q=-p$, the intersection plane is not unique, and any tangent vector $X\in \tspace{p}{\Sn{2}}$ with $\lVert X \rVert_2 = \pi$ is a solution of the above construction.

Vectors tangent to the sphere at different points cannot be added or subtracted.
First, they need to be transported to the same point.
This is performed using parallel transport.
Transport of a vector $X$ tangent at $p$ to a point $q$ is denoted $\partrans{p}{q} X$.
This operation results in a new vector tangent to the sphere at $q$.
Note that transporting a vector $X$ tangent at $p$ between a series of points -- for example from $p$ to $q$, from $q$ to $r$, and from $r$ back to $p$ -- may not result in the same vector $X$.

These operations can be extended to all manifolds as described in  Section~\ref{section:general-definitions}.

\subsection{Example: two metrics on the symmetric positive-definite matrices}
\label{section:example-spd}

A second, more elaborate example is the set of $n\times n$ symmetric positive definite (SPD) matrices $\symposdef{n}$.
The set consists of all matrices $p \in \Rn{n\times n}$ that are symmetric ($p = p^\tT$) and positive definite,
that is, for any $0\neq x \in \Rn{n}$, it holds that $x^\tT px > 0$, which equivalently means that all eigenvalues of $p$ are strictly positive.

The “directions to walk in”, or to be precise, the tangent space $T_p\symposdef{n}$ is the set of symmetric matrices $X=X^\tT$.
Note that for an arbitrary symmetric matrix $q = p + tX$ where $t>0$, $q$ might only be SPD for very small values of $t$.
Similar to the sphere, addition might mean leaving the manifold.

There are several metrics we can introduce on the tangent spaces $T_p\symposdef{n}$. In the following we compare two and their effect on the exponential map, or informally said “how to move into a direction $X$”.
One possible metric is the affine invariant metric, given by~\cite{pennec_riemannian_2019}
\begin{equation*}
    \ip{X}{Y}_{p,\mathrm{AI}} = \operatorname{tr}(p^{-1}Xp^{-1}Y).
\end{equation*}
With this metric, the exponential map is
\begin{equation*}
    \exp_p X = p^{\frac{1}{2}}\operatorname{Exp}(p^{-\frac{1}{2}} X p^{-\frac{1}{2}})p^{\frac{1}{2}},
\end{equation*}
where $\operatorname{Exp}$ is the matrix exponential.
The simplest case of $n=1$ yields the manifold of positive numbers, whose inner product simplifies to $\ip{X}{Y}_{p,\mathrm{AI}} = \frac{XY}{p^2}$ and whose exponential is $\exp_pX = p\mathrm{e}^{X/p}$.
Hence, non-positive numbers can never be reached.
The main computational effort for this metric is the computation of square roots and exponentials.

Another possibility is to use the Bures-Wasserstein metric, which is given by \cite{BHATIA2019165}
\begin{equation*}
    \ip{X}{Y}_{p,\mathrm{BW}} = \frac{1}{2}\operatorname{tr}(L_p(X)Y),
\end{equation*}
where $q=L_p(X)$ denotes the Lyapunov operator, which solves $pq + qp = X$.
The exponential map then reads
\begin{equation*}
    \exp_p X = p+X+L_p(X)pL_p(X),
\end{equation*}
where for both operations the computational effort is mainly to compute the Lyapunov operator.

\section{Structure of the library}
\label{section:structure-of-library}

\lstinline{Manifolds.jl}\footnote{The documentation for \lstinline!Manifolds.jl! is available at \url{https://juliamanifolds.github.io/Manifolds.jl/stable/}.}
is a registered Julia package. Its most recent version, currently 0.8.61, can be easily installed typing
\lstinline!using Pkg; Pkg.add("Manifolds")! in the Julia REPL and activated by typing \lstinline!using Manifolds! afterwards.

The library of manifolds \lstinline!Manifolds.jl! is based on the interface \lstinline!ManifoldsBase.jl!\footnote{The documentation for \lstinline!ManifoldsBase.jl! is available at \url{https://juliamanifolds.github.io/ManifoldsBase.jl/stable/}.},
whose current version is 0.14.5.
This interface defines a common base for Riemannian manifolds in Julia.
It hence serves two purposes. On one hand it can be used to implement additional manifolds like the ones in \lstinline!Manifolds.jl!.
On the other hand it can be used to define arbitrary functionality on arbitrary manifolds without depending on or loading the whole library of manifolds.

Both the library and the interface use GitHub Actions to automatically generate their documentation as well as run tests, following the philosophy of continuous integration. These tests are run on Linux and Mac OS with the latest (Julia 1.9), long-term-support (Julia 1.6), and nightly versions of Julia.
For pull requests (PRs) and for commits on the master branch a format test is run to ensure that all source code follows the formatting rules using \lstinline!JuliaFormatter.jl!%
\footnote{See \url{https://juliamanifolds.github.io/Manifolds.jl/stable/misc/contributing.html} for details.}

Using the \lstinline!JuliaRegistrator! GitHub application and \lstinline!TagBot! GitHub action, new releases are automatically registered to the registry and new GitHub release entries as well as documentation page generation is triggered.

Currently, the code coverage, \ie percentage of code lines executed during testing, is 99.86\,\% for \lstinline!ManifoldsBase.jl! and 98.93\,\% for \lstinline!Manifolds.jl!, respectively.

\subsection{Manifolds and functions thereon}

A Riemannian manifold in \lstinline!Manifolds.jl! is defined by introducing a subtype of \lstinline!AbstractManifold{𝔽}!, where $\mathbb F$  is the field the manifold is based on: the real numbers $\R$, complex numbers $\mathbb{C}$, or quaternions $\mathbb{H}$. For example,
\begin{jllisting}[language=Julia]
M1 = Sphere(2)
M2 = Euclidean(2, 3, ℂ)
M3 = SymmetricPositiveDefinite(3)
\end{jllisting}
creates the sphere $\Sn{2} \subset \Rn{3}$ from \ref{section:example-sphere}, the space of complex-valued $2\times 3$ matrices $\mathbb C^{2\times 3}$, and the set of $3\times 3$ symmetric positive definite matrices $\symposdef{3}$ from \ref{section:example-spd}, respectively.

The interface provides a unified way to work with data defined on arbitrary manifolds, including those listed above.
As a first example, \lstinline!manifold_dimension(M)! returns the dimension of the manifold.
The two functions \lstinline!is_point(M, p)! and \lstinline!is_vector(M, p, X)! can be used to check whether points or tangent vectors are valid.
The last two have an optional positional parameter to activate throwing an error for invalid points or tangent vectors, which is \lstinline!false! by default.
Hence calling \lstinline!is_point(M, p, true)! returns \lstinline!true! if \lstinline!p! is a point on \lstinline!M! and otherwise throws an error describing why \lstinline!p! is not a point on \lstinline!M!.

For functions that return points or tangent vectors, interface functions have both in-place and allocating versions.
By default, the allocating version allocates an output and then calls the in-place version.
For example, for the exponential map, the function \lstinline{exp!(M, q, p, X)} stores $\exp_pX$ in the variable \lstinline!q!.
When the in-place variant is implemented for a new manifold, the allocating version \lstinline!exp(M, p, X)! is automatically available.
Calling \lstinline!exp(M, p, X)! first allocates memory and then calls the in-place function mentioned above.
The same holds for \lstinline!geodesic(M, p, X)!, which returns the function $\gamma_{p,X}$.
To evaluate the geodesic at a certain $t$ one can call \lstinline!geodesic(M, p, X, t)!.
In both cases, in-place variants are available as well.
The same principle of in-place and allocating variants is also realized for \lstinline!log(M, p, q)! and \lstinline{log!(M, X, p, q)}, which should return a deterministic result if the logarithmic map is not unique as for opposite points on the sphere.
With the latter, the former is available with a default implementation as well as \lstinline!shortest_geodesic(M, p, q)! (and its in-place variant) again with \lstinline!t! as an optional further parameter, similar to \lstinline!geodesic!.
While all these default implementations are available, they can be overwritten if for example the shortest geodesic $\gamma_{p;q}$ can be computed more efficiently than by calling $\exp_p(t\log_pq)$.
Finally, the parallel transport $\partrans{p}{q} X$ is implemented as \lstinline!parallel_transport_to(M, p, X, q)! and \lstinline{parallel_transport_to!(M, Y, p, X, q)} in place of \lstinline!Y!, respectively.

When using \lstinline!SymmetricPositiveDefinite(n)!, the affine-invariant metric is used by default.

\subsection{Representation of points and vectors}

Usually points $p\in\cM$ and tangent vectors $X\in\tspace{p}{\cM}$ are not explicitly typed, and only the manifolds type is specified for dispatch.
Most manifolds implicitly assume that both points and tangent vectors are instances of \lstinline! AbstractArray! or at least have several functions like addition defined.
This enables a great deal of flexibility in types to use for these parameters in a function, such as the statically sized arrays from \lstinline!StaticArrays.jl!, which for small dimensions are stack-allocated.
In two situations the points and tangent vectors are typed.
If the representation requires more than one array, as is the case for the fixed-rank matrices, then a type is used to enforce this.
Since the interface functions are not typed, features like the allocating variant of a function being available as soon as the in-place one is implemented still work.
Another reason for using types is to distinguish different representations of points and tangent vectors on one manifold.
For example, for the hyperbolic space $\Hn{n}$, three representations are implemented:
the hyperboloid model, the Poincaré ball model, and the Poincaré half-space model.
These are distinguished by using different types for points and tangent vectors, such as
\lstinline!PoincareBallPoint! and \lstinline!PoincareBallTVector! for points and tangent vectors in the Poincaré ball model.
Using (plain) arrays on $\Hn{n}$ is equivalent to using \lstinline!HyperboloidPoint! and \lstinline!HyperboloidTVector!.

\subsection{Retractions and vector transports}

The exponential map $\exp_p$, the logarithmic map $\log_p$, and parallel transport $\partrans{p}{q}$ are computationally expensive or not given in closed form for some manifolds.
For these cases one can specify subtypes of \lstinline!AbstractRetractionMethod!, \lstinline!AbstractInverseRetractionMethod!, or \lstinline!AbstractVectorTransportMethod! and create objects of these subtypes, respectively denoted \lstinline!r!, \lstinline!s!, and \lstinline!v! in the following example.
The corresponding functions are called by \lstinline!retract(M, p, X, r)!, \lstinline!inverse_retract(M, p, q, s)!, and \lstinline!vector_transport_to(M, p, X, q, v)!.
These methods also have in-place variants.
On the sphere $\Sn{2}$, for example, \lstinline!r = ProjectionRetraction()! computes the retraction given by projecting $p+X$ onto the sphere.

These three methods can be used to implement generic algorithms on manifolds.
Since the exponential map might not be given in closed form, the interface
provides the function \lstinline!default_retraction_method(M)! to obtain the retraction that is most often used on a certain manifold.
If the exponential map is available (and can be computed efficiently), this method returns \lstinline!ExponentialRetraction()!, which lets \lstinline!retract! fall back to using \lstinline!exp!.
This way algorithms can easily be modified to use different available retractions.
The same scheme applies for inverse retractions and vector transports.

\subsection{Features on manifolds using a decorator pattern}

As mentioned above, the manifold \lstinline!M3 = SymmetricPositiveDefinite(3)! implicitly uses the affine invariant metric. To change that, the interface \lstinline!ManifoldsBase.jl! provides a trait-based decorator pattern. To decorate the manifold with the Bures-Wasserstein metric, we can declare
\begin{jllisting}[language=Julia]
M3b = MetricManifold(M3, BuresWassersteinMetric())
\end{jllisting}
This acts like a decorator for the manifold in the following sense: all methods that depend on the metric, like \lstinline!exp! and \lstinline!log! mentioned above or \lstinline!inner(M3b, p, X, Y)! and \lstinline!norm(M3b, p, X)!, are implemented differently for this manifold.
Other methods like \lstinline!manifold_dimension! or \lstinline!is_point! that do not depend on the metric are “passed down” to \lstinline!M3! automatically.

Traits can be seen as a list of properties “attached” to a manifold.
For example, \lstinline!M2 = Euclidean(2, 3, ℂ)! has the property \lstinline!IsDefaultMetric(EuclideanMetric())!, that is, using the manifold itself is completely equivalent to explicitly equipping the manifold with the same metric \lstinline!M2b = MetricManifold(M2, EuclideanMetric())!.
This formulation allows a manifold to be implemented with an implicit default metric without needing to directly implement a metric type or use \lstinline!MetricManifold!.
But one can still implement a second metric and decorate the manifold with it.

Similarly, a default embedding can be implicitly assumed, for example by implementing \lstinline!get_embedding(M)!, \lstinline!embed(M, p)! and \lstinline!project(M, p)!.
An embedding of \lstinline!M! into a different manifold \lstinline!N! can be represented by \lstinline!EmbeddedManifold(M, N)! similarly to the metric above.
For example, for the sphere \lstinline!M1 = Sphere(2)! we have that \lstinline!get_embedding(M1)! returns \lstinline!Euclidean(3)!, and \lstinline!embed(M1, p)! is the identity function since points are represented by unit vectors.
The projection \lstinline!project(M1, p)! normalises \lstinline!p! to norm 1.
Also, for a Lie group, this implicit approach can be realized by implementing \lstinline!compose(M, p, q)!.
The explicit decorator is the \lstinline!GroupManifold(M, op)!, where \lstinline!op! is the Lie group operation.

As a final example of a decorator, we mention the \lstinline!ValidationManifold!. For all default functions mentioned above, this manifold adds checks (\lstinline!is_point! and \lstinline!is_vector!) to their input and output values. While this somewhat reduces performance, using \lstinline!M1b = ValidationManifold(M1)! provides an easy way to verify that none of the operations receive invalid inputs or produce invalid outputs.
Here the \lstinline!IsExplicitDecorator! property is set; that is, if no new method is implemented for the validation manifold \lstinline!M1b!, the method for \lstinline!M1! is automatically called.

When implementing a new manifold, the properties or traits are set using the function \lstinline!active_traits!. For example, for the sphere this is defined as
\begin{jllisting}
function active_traits(f, ::AbstractSphere, args...)
    return merge_traits(IsIsometricEmbeddedManifold(), IsDefaultMetric(EuclideanMetric()))
end
\end{jllisting}
which finally illustrates that setting \lstinline!IsIsometricEmbeddedManifold()! specifies that the metric from the embedding is used and does not need to be implemented.
Here the metric $\ip{X}{Y}_p$ is computed when calling \lstinline!inner(M1, p, X, Y)! as \lstinline!inner(get_embedding(M1), embed(M1, p), embed(M1, p, X), embed(M1, p, Y))!.
It does not need its own implementation.
Since both embedding points and embedding tangent vectors are just the identity function for the sphere, this does not even introduce any overhead.
Further, since all the information required for these dispatches is given by the type of the manifold, this dispatch-logic can be determined at compile time.
It therefore does not yield any runtime overhead.

\section{Comparison to other libraries}
\label{sec:related-work}

Table~\ref{tab:manifolds-of-libraries} compares availability of manifolds in \lstinline!Manifolds.jl! 0.8.61 to other existing libraries: Geomstats 2.5.0 \citep{miolane_geomstats_2020}, Geoopt 0.5.0 \citep{kochurov_geoopt_2020}, Manopt 7.1 \citep{manopt}, McTorch 1.1 \citep{meghwanshi_mctorch_2018}, Pymanopt 2.1.1 \citep{townsend_pymanopt:_2016}, ROPTLIB 0.8 \citep{huang_roptlib_2018}, TheanoGeometry \citep{kuhnel_computational_2017}, Jax Geometry~\citep{kuhnel_computational_2017}, and TensorFlow RiemOpt 0.1.2~\citep{Smirnov:2021}.
Availability of different operations is compared in Table~\ref{tab:features-of-libraries}.

\begin{table}[tbp]
    \vspace*{-3.5\baselineskip}
    \centering
    \caption{\small Comparison of manifolds available in related libraries.
    A manifold is marked as available if at least a retraction is implemented using a generic interface.
    The fields \fR, \fC, \fH\ are meant in increasing order, \fN indicating that a manifold is not available.\\
    $^{\mathrm{a}}$ this manifold is available for arrays with any number of indices.
    \\
    $^{\mathrm{b}}$ only the two-dimensional manifold is available.
    \\[-1.9\baselineskip]
    }
    \label{tab:manifolds-of-libraries}
    \footnotesize
    \begin{tabular}{@{}L{5.7cm}C{0.6cm}C{0.6cm}C{0.6cm}C{0.6cm}C{0.6cm}C{0.6cm}C{0.6cm}C{0.6cm}C{0.6cm}L{1.5cm}@{}}
    \toprule
    Manifold & \tHead{Geomstats} & \tHead{Geoopt} & \tHead{Manifolds.jl} & \tHead{Manopt} & \tHead{McTorch} & \tHead{Pymanopt} & \tHead{ROPTLIB} & \tHead{TheanoGeometry} & \tHead{Jax Geometry} & \tHead{TF\,RiemOpt} \\ \midrule
    Centered matrices     & \fN & \fN & \fC & \fR & \fN & \fN & \fN & \fR & \fN & \fN \\
    Discretized curves    & \fR & \fN & \fN & \fN & \fN & \fN & \fN & \fN & \fN & \fN \\
    Ellipsoid             & \fN & \fN & \fN & \fN & \fN & \fN & \fN & \fR & \fR & \fN \\
    Elliptope             & \fN & \fN & \fR & \fR & \fN & \fN & \fN & \fN & \fN & \fN \\
    Embedded cylinder     & \fN & \fN & \fN & \fN & \fN & \fN & \fN & \fN & \fR & \fN \\
    Essential manifold    & \fN & \fN & \fR & \fR & \fN & \fN & \fN & \fN & \fN & \fN \\
    Euclidean             & \fR & \fR & \fH[\mathrm{a}] & \fC & \fR & \fR & \fR & \fR & \fN & \fR \\
    Fixed rank matrices   & \fN & \fN & \fC & \fR & \fN & \fN & \fC & \fN & \fN & \fN \\
    Flag           & \fN & \fN & \fR & \fN & \fN & \fN & \fN & \fN & \fN & \fN \\
    General linear group  & \fR & \fN & \fC & \fN & \fN & \fN & \fN & \fR & \fN & \fN \\\midrule
    General multin.~doubly stoc.~matrices
                          & \fN & \fN & \fN & \fR & \fN & \fN & \fN & \fN & \fN & \fN \\
    Generalized Grassmann & \fN & \fN & \fC & \fR & \fN & \fN & \fN & \fN & \fN & \fN \\
    Generalized Stiefel   & \fN & \fN & \fC & \fR & \fN & \fN & \fN & \fN & \fN & \fN \\
    Grassmann             & \fR & \fN & \fC & \fC & \fN & \fR & \fR & \fN & \fN & \fR \\
    Heisenberg group      & \fR & \fN & \fR & \fN & \fN & \fN & \fN & \fN & \fR & \fN \\
    Hilbert sphere        & \fN & \fN & \fN & \fN & \fN & \fN & \fR & \fN & \fN & \fN \\
    Hyperbolic space      & \fR & \fR & \fR & \fR & \fN & \fN & \fN & \fN & \fR[\mathrm{b}] & \fR \\
    Kendall's preshape space & \fR & \fR & \fR & \fR & \fN & \fR & \fN & \fR & \fR & \fN \\
    Kendall's shape space & \fR & \fN & \fR & \fN & \fN & \fN & \fN & \fR & \fR & \fN \\
    Lorentzian manifold   & \fR & \fR & \fR & \fN & \fN & \fN & \fN & \fN & \fN & \fN \\\midrule
    Multin.~doubly stoc.~matrices
                          & \fN & \fR & \fR & \fR & \fN & \fN & \fN & \fN & \fN & \fN \\
    Multinomial matrices  & \fN & \fN & \fR & \fR & \fN & \fN & \fN & \fN & \fN & \fN \\
    Multinomial symmetric matrices
                          & \fN & \fN & \fR & \fR & \fN & \fN & \fN & \fN & \fN & \fN \\
    Oblique manifold      & \fN & \fN & \fR & \fR & \fN & \fR & \fR & \fN & \fN & \fN \\
    Phases of real DFT    & \fN & \fN & \fN & \fC & \fN & \fN & \fN & \fN & \fN & \fN \\
    Poincaré polydisk     & \fR & \fN & \fN & \fN & \fN & \fN & \fN & \fN & \fN & \fN \\
    Positive numbers      & \fN & \fN & \fR & \fR & \fN & \fR & \fR & \fN & \fN & \fN \\
    Probability simplex   & \fR & \fN & \fR & \fR & \fN & \fN & \fN & \fN & \fN & \fN \\
    Product Space         & \fR & \fR & \fH & \fC & \fN & \fC & \fC & \fN & \fN & \fR \\
    Projective Space      & \fR & \fN & \fH[\mathrm{a}] & \fC & \fN & \fR & \fR & \fN & \fN & \fR  \\\midrule
    Rotations             & \fR & \fN & \fR & \fR & \fN & \fR & \fR & \fR & \fN & \fR \\
    Skew-symmetric matrices
                          & \fR & \fN & \fC & \fR & \fN & \fN & \fN & \fN & \fN & \fN \\
    Special Euclidean group
                          & \fR & \fN & \fR & \fR & \fN & \fN & \fN & \fN & \fN & \fN \\
    Spectrahedron         & \fN & \fN & \fR & \fR & \fN & \fN & \fN & \fN & \fN & \fN \\
    Sphere / Circle       & \fR & \fR[\mathrm{a}] & \fC[\mathrm{a}] & \fC[\mathrm{a}] & \fR & \fC[\mathrm{a}] & \fC & \fR & \fR[\mathrm{b}] & \fR \\
    Stiefel               & \fR & \fR & \fC & \fC & \fR & \fR & \fC & \fN & \fN & \fR \\
    Sym. matrices    & \fR & \fN & \fC & \fR & \fN & \fN & \fN & \fN & \fN & \fN \\
    Sym. pos. def. matrices
                          & \fR & \fN & \fR & \fR & \fR & \fR & \fR & \fR & \fN & \fR  \\
    Sym. pos. def. simplex
                          & \fN & \fN & \fN & \fC & \fN & \fN & \fN & \fN & \fN & \fN \\
    Sym. pos. semidef. fixed rank matrices
                          & \fR & \fN & \fC & \fC & \fN & \fC & \fC & \fN & \fN & \fN \\\midrule
    Symplectic            & \fN & \fN & \fR & \fN & \fN & \fN & \fN & \fN & \fN & \fN \\
    Symplectic Stiefel    & \fN & \fN & \fR & \fN & \fN & \fN & \fN & \fN & \fN & \fN \\
    Tangent bundle with Sasaki metric & \fR & \fN & \fR & \fN & \fN & \fN & \fN & \fN & \fN & \fN \\
    Tensors of fixed tensor train rank & \fN & \fN & \fN & \fR & \fN & \fN & \fN & \fN & \fN & \fN \\
    Tucker manifold & \fN & \fN & \fC & \fR & \fN & \fN & \fN & \fN & \fN & \fN \\
    Torus           & \fR & \fR & \fR & \fR & \fN & \fR & \fR & \fN & \fR & \fR \\
    Unit-norm sym. matrices
                          & \fN & \fN & \fC & \fR & \fN & \fN & \fN & \fN & \fN & \fN  \\ \bottomrule
    \end{tabular}
\end{table}

Note that the extent of support varies significantly across libraries.
For example \lstinline!Manifolds.jl! currently implements five different metrics (affine-invariant, Bures-Wasserstein, Log-Euclidean,  Log-Cholesky, and generalized Bures-Wasserstein) on the manifold of symmetric positive definite matrices, Geomstats only the first three ones and a Power-Euclidean one, Manopt only the first two, and Pymanopt only the first.

On the other hand, \lstinline!Manifolds.jl! focuses solely on implementing Riemannian manifolds, metrics, Lie groups, and functions thereon. Geomstats further provides several statistical tools, and Manopt and Pymanopt further focus on optimisation algorithms, such as a gradient descent scheme. For optimisation on manifolds in Julia, we refer to \lstinline!Manopt.jl! \citep{Bergmann:2022}, which, like Manopt and Pymanopt, focuses also on nonsmooth optimisation algorithms on manifolds. For that reason, Table~\ref{tab:manifolds-of-libraries} compares features provided by the libraries that focus on implementations of manifolds but not statistics and/or optimisation.

All packages but ROPTLIB provide support for several Automatic Differentiation (AD) backends to compute Euclidean gradients and Hessians of functions defined in the embedding of a manifold and to convert them to Riemannian variants afterwards.

\begin{table}[tbp]
  \centering
  \caption{Comparison of features available in related libraries.}
  \label{tab:features-of-libraries}
    \footnotesize
    \begin{tabular}{@{}L{5.2cm}C{0.6cm}C{0.6cm}C{0.6cm}C{0.6cm}C{0.6cm}C{0.6cm}C{0.6cm}C{0.6cm}C{0.6cm}L{1.5cm}@{}}
    \toprule
    Feature & \tHead{Geomstats} & \tHead{Geoopt} & \tHead{Manifolds.jl} & \tHead{Manopt} & \tHead{McTorch} & \tHead{Pymanopt} & \tHead{ROPTLIB} & \tHead{TheanoGeometry} & \tHead{Jax Geometry} & \tHead{TF\,RiemOpt}  \\ \midrule
  Atlases and charts          & \fM & \fM & \fP & \fM & \fM & \fM & \fM & \fM & \fP & \fM \\
  Automatic differentiation   & \fP & \fP & \fP & \fP & \fP & \fP & \fM & \fP & \fP & \fP  \\
  Compatibility with extended precision floating point types
                              & \fM & \fM & \fP & \fP & \fM & \fM & \fM & \fM & \fM & \fM \\
  Bases of tangent spaces     & \fM & \fM & \fP & \fP & \fM & \fM & \fM & \fM & \fM & \fM \\\midrule
  Connections                 & \fP & \fM & \fP & \fM & \fM & \fM & \fM & \fM & \fM & \fM \\
  Exponential maps (retractions)
                              & \fP & \fP & \fP & \fP & \fP & \fP & \fP & \fP & \fP & \fP \\
  Inner product               & \fP & \fP & \fP & \fP & \fP & \fP & \fP & \fP & \fP & \fP \\
  Lie groups and actions      & \fP & \fM & \fP & \fM & \fM & \fM & \fM & \fP & \fN & \fN  \\\midrule
  Logarithmic maps (inverse retractions)
                              & \fP & \fP & \fP & \fP & \fP & \fP & \fP & \fP & \fP & \fP \\
  Riemannian gradients        & \fM & \fP & \fP & \fP & \fP & \fP & \fP & \fP & \fM & \fP \\
  Riemannian Hessians         & \fM & \fP & \fM & \fP & \fP & \fP & \fP & \fM & \fM & \fM \\
  Vector transport            & \fP & \fP & \fP & \fP & \fP & \fP & \fP & \fP & \fP & \fP \\
  \bottomrule
  \end{tabular}
  \end{table}

\section{Examples}
\label{sec:examples}

This section demonstrates a few examples of composing generic operations from \lstinline!Manifolds.jl! with the Julia ecosystem~\citep{bezanson_julia:_2017}.
The first example defines a Bézier spline on a generic manifold to demonstrate how to easily define new functions on manifolds using \lstinline!ManifoldsBase.jl!.
Then, \lstinline!Manopt.jl!\footnote{Current version 0.3.39, available at \url{https://manoptjl.org}} \citep{Bergmann:2022} is used as an example of a complex optimization library built on top of \lstinline!Manifolds.jl!.
Further, it is demonstrated how \lstinline!Manifolds.jl! can be easily combined with independently developed libraries to implement fast, geometry-aware algorithms.

Note that all examples can be split into two components.
First, a manifold and data on the manifold are constructed.
Second, the algorithm is executed on this manifold.
Since all methods are implemented in a generic way, switching to data from another manifold just requires to change the first part of introducing the manifold and loading/generating corresponding data.

Each of the examples illustrates one feature original to Manifolds.jl. While all of them are implemented agnostic of a specific manifold at hand, using a recursion by dispatch is unique to Julia. An optimisation problem can be set up in three lines, defining the cost and the gradient as inline functions. The last example illustrates the generic implementation of bases of tangent spaces. These can even be abstract in the sense that no tangent space basis vectors are stored and they are only allocated if needed.

\subsection{Implementing functions on manifolds}
\label{sec:fun-on-manifolds}

Functions on manifolds dispatch on the manifold itself, which is included as the first argument.
As an example, we implement a Bézier curve on an arbitrary manifold.
Bézier curves are used to model curves and surfaces in computer graphics.
Their shape is determined by so called control points.
The curve starts in the first control point and ends in the last.
For more details on their use on manifolds, see for example \citep{AbsilGousenbourgerStriewskiWirth:2016, GousenbourgerMassartAbsil:2018, BergmannGousenburger-2018}.

A Bézier curve can be defined using the De Casteljau algorithm.
Let us first consider the classical Euclidean case: given a degree $n$ and $n+1$ (control) points $x_0,\ldots,x_n\in\Rn{n}$, the Bézier curve $b_n(t; x_0,\ldots,x_n)$ is defined recursively as
\begin{align*}
    b_n(t; x_0,\ldots,x_n)
    &=
    b_1(t; b_{n-1}(t; x_0,\ldots,x_{n-1}), b_{n-1}(t;x_1,\ldots,x_n))
    \\
    b_1(t; x_0,x_1) &= x_0 + t(x_1-x_0)
\end{align*}

A Bézier curve $b_1(t; x_0,x_1)$ of order $1$ is just a line.
To evaluate a Bézier curve $b_2(t; x_0,x_1,x_2)$ at a point $t$, we would need three line evaluations: $y_0=b_1(t;x_0,x_1)$, $y_1=b_1(t;x_1,x_2)$ and their connecting line $b_2(t;x_0,x_1,x_2) = b_1(t;y_0,y_1)$.
The result is a quadratic curve in $\Rn{n}$.

On a Riemannian manifold $\cM$ we obtain a Bézier curve by evaluating shortest geodesics $\gamma(t; p,q)$ connecting $p,q\in\cM$ instead of line segments.

The implementation for an arbitrary Riemannian manifold is given by the following code and returns the Bézier curve evaluated at $t$.

\jlinputlisting{bezier.jl}

\begin{figure}\centering
    \includegraphics[width=.5\textwidth]{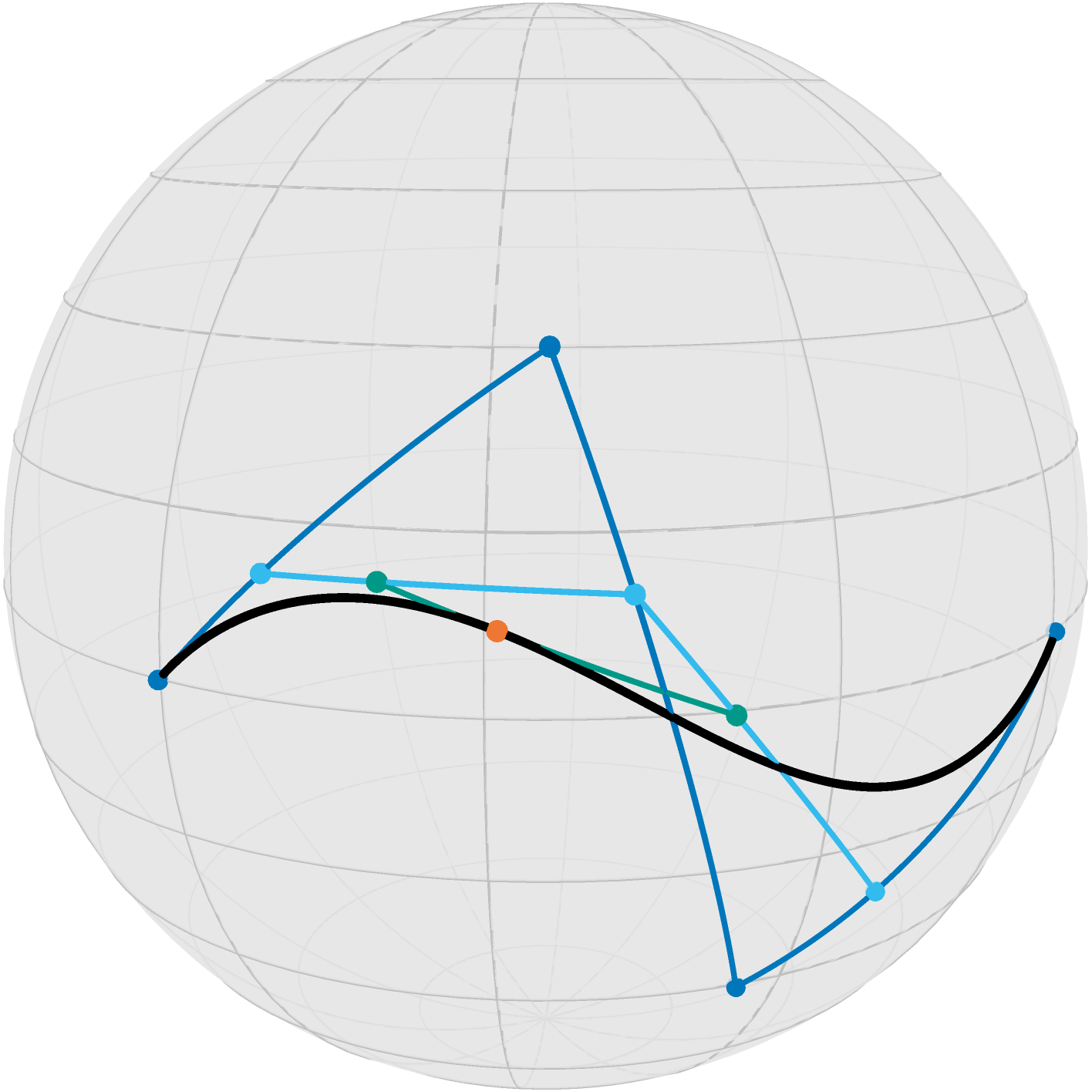}
    \caption{
        Illustration of the De Casteljau algorithm to evaluate a Bézier curve on the Sphere $\Sn{2}$.
        The initial points and their shortest geodesics (blue) yield three points (cyan) and two points (green) during the recursion, and their connecting geodesic the point on the Bézier curve (orange).
        }
    \label{fig:Bezier}
\end{figure}

For four points on the sphere given by
\begin{equation*}
p_0 = \begin{pmatrix}
    0&-1&0\end{pmatrix}^{\mathrm{T}}\!\!,
\quad
p_1 = \begin{pmatrix}
    -\frac{1}{2}&-\frac{1}{\sqrt{2}}&-\frac{1}{2}
\end{pmatrix}^{\mathrm{T}}\!\!,
\quad
p_2 = \begin{pmatrix}
    -\frac{1}{\sqrt{2}}&-\frac{1}{2}&\frac{1}{2}
\end{pmatrix}^{\mathrm{T}}\!\!,
\quad
p_3 = \begin{pmatrix}
-1&0&0
\end{pmatrix}^{\mathrm{T}}\!\!,
\end{equation*}
and an evaluation of $b_3$ at $t=\frac{2}{3}$ all shortest geodesics involved in the evaluation are shown in Figure \ref{fig:Bezier}.
This figure has been adapted from \citep[Figure 9]{BergmannGousenburger-2018}.

\subsection{Optimization using Manopt.jl}
\label{sec:manopt-jl}

Now we consider the generalization of the mean to Riemannian manifolds.
The mean of some (Euclidean) data $x_1,\ldots,x_N\in\Rn{n}$ can be computed as
\begin{equation*}
    \hat x = \displaystyle\frac{1}{N}\sum_{k=1}^N x_k.
\end{equation*}
One can interpret this formula as the solution of the first-order optimality condition of the optimization problem
\begin{equation*}
    \hat x = \operatorname*{arg\,min}_{y\in\Rn{n}}\sum_{k=1}^N\lVert x_k-y\rVert^2.
\end{equation*}
That is, $\hat x$ is the point $y$ that minimizes the sum of squared distances, or equivalently, the variance.
This interpretation can be generalized to manifolds, but usually no closed form for the minimizer $\hat x$ exists.
Still, replacing the squared distance with the squared Riemannian distance, the gradient can easily be computed to perform a gradient descent algorithm with a suitable line search strategy, cf.~for example~\cite[Chapter 4]{AbsilMahonySepulchre:2008}.
For given points $p_1,\ldots,p_N\in\cM$ the cost function and the gradient read
\begin{equation*}
    F(q) = \frac{1}{N}\sum_{k=1}^N d_{\cM}^2(q,p_k)
    \quad\text{and}\quad
    \operatorname{grad}F(q) = -\frac{1}{2N}\sum_{k=1}^N \log_q p_k.
\end{equation*}
This algorithm \lstinline!gradient_descent! is implemented in \lstinline!Manopt.jl!. Furthermore, a library of cost functions, gradients, differentials, adjoint differentials, as well as proximal maps is available in the package as well. They are implemented on arbitrary manifolds based only on \lstinline!ManifoldsBase.jl! and can hence be used to design an optimization problem even independent of the manifold.
One example that we also require here is the already mentioned gradient of the (exponentiated) distance function \lstinline!distance(M,p,q)^s! with respect to \lstinline!q!, where \lstinline!s=2! is the default. This function is called \lstinline!grad_distance(M, p, q, s)! in \lstinline!Manopt.jl!.

Note that since \lstinline!Manopt.jl! is based on the lightweight interface \lstinline!ManifoldsBase.jl!, we need to load \lstinline!Manifolds.jl! only to have our manifold of interest available.

\jlinputlisting{mean_manopt.jl}

\lstinline!Manifolds.jl! itself includes two approaches for computing the weighted mean: gradient descent and on-line repeated geodesic interpolation \citep{HoChengSalehianVemuri_2013}.
Using these methods, implementations of weighted variance, skewness, kurtosis, and higher moments are also provided.

\subsection{Tangent space PCA}
\label{sec:tpca}

Thanks to the high composability of the Julia ecosystem and the design of the \lstinline!Manifolds.jl! library, it is very easy to compose independent libraries to achieve new functionality.
For example, tangent space principal component analysis (PCA) described by~\cite{fletcher_principal_2004} can be split into 1) computing the Riemannian centers of mass and coordinates of tangent vectors using \lstinline!Manifolds.jl! and 2) calculation of PCA vectors using \lstinline!MultivariateStats.jl!.
Using a $d$-dimensional manifold, tangent vectors might be stored in a format different from the $d$-dimensional vector. The interface \lstinline!ManifoldsBase.jl! provides \lstinline!c = get_coordinates(M, p, X, B)! to obtain a $d$-dimensional representation and \lstinline!X = get_vector(M, p, x, B)! to reconstruct the tangent space format. For example tangent vectors from $\tspace{p}{\Sn{2}}$ are stored as vectors $X$ orthogonal to $p$ (in $\Rn{3}$) and they have a representation in 2 coordinates.
Both these functions require a basis \lstinline!B!, which can be a full set of vectors in a \lstinline!CachedBasis! or a memory-efficient type to represent a deterministic function of a \lstinline!DefaultOrthonormalBasis! to not store the basis explicitly.
The whole procedure is demonstrated by the following code.

\jlinputlisting{tangent_space_pca.jl}

\section{Benchmarks}
\label{sec:benchmarks}

This section presents and discusses benchmarks that compare performance of \lstinline!Manifolds.jl! to other libraries.
These are Geomstats~\cite{miolane_geomstats_2020} with its backends AutoGrad, NumPy, PyTorch, and TensorFlow;
Geoopt~\citep{kochurov_geoopt_2020}, which is based on PyTorch; Pymanopt~\cite{townsend_pymanopt:_2016}; TensorFlow RiemOpt~\cite{Smirnov:2021}; Manopt~\cite{manopt} in Matlab; and ROPTLIB~\cite{huang_roptlib_2018} as a C++ library.
We build three groups of manifolds: The Euclidean space $\Rn{n}$, the hyperbolic space $\Hn{n}$, and the sphere $\Sn{n}$ for the dimensions $n=2, 3, 4, 8, 16, 32, 2^{10}, 2^{15}, 2^{20}$.
For these, ROPTLIB only provides an exponential map; the hyperbolic manifold is missing for ROPTLIB and Pymanopt.

The second set of manifolds are the special orthogonal group~$\operatorname{SO}(n)$,
the symmetric positive definite matrices $\mathcal P(n)$ with the affine invariant metric, and for a high dimensional test the power manifold $(\mathcal P(n))^{128\times 128}$, where in all three cases we used
$n=2,3,4,8,16,32$.
Here again, ROPTLIB only provides an exponential map for the last two manifolds.
The power manifold as well as a distance on $\mathrm{SO}(n)$ is not available in Geomstats and GeoOpt.

\subsection{General setup}

A computer running Linux Mint 20 Ulyana with Intel(R) Core(TM) i7-9700KF CPU @ 3.60GHz CPU and 32 GB of RAM was used to benchmark all libraries.
In particular, no limitations were imposed on the number of threads.

The benchmarking was performed using points generated randomly by \lstinline!Manifolds.jl!.
Each benchmarked operation was performed using 100 pairs of these points $10^N$ times in a loop written in the language native to the library.
The number $N$ was chosen to be large enough to ensure that the entire benchmarking loop takes at least one tenth of a second to reduce the influence of system clock inaccuracy.
Results of operations were accumulated in an array, which incurs a small performance penalty but prevents the compiler from optimizing out an otherwise unused computation.
Details of benchmarks specific to particular libraries and the settings used are given in Appendix~\ref{app:TechnicalDetails}.

\subsection{Discussion}

The results for the first experiment are shown in Figure~\ref{fig:benchmark}.
On all three manifolds, \lstinline!Manifolds.jl! performs the fastest for the smaller dimensions $n=2,3,4,8,16$, and for example in the sphere for $n=3$ ROPTLIB performs the exponential map (bottom center bar plot) only a factor of $2$ slower.
This is likely due to the more sophisticated memory management and use of specialized linear algebra packages in \lstinline!Manifolds.jl!.
\\
For larger dimensions, e.\,g.\ for the manifold $\mathcal M = \Rn{2^{10}}$, Pymanopt computes the logarithmic map about 1/6 faster. Similarly for $n=32$, ROPTLIB computes the exponential map on the sphere $\Sn{n}$ about 33\% faster than \lstinline!Manifolds.jl! and on $\Rn{n}$ about four times faster.
\\
For very high dimensions, RiemOpt with its TensorFlow backend is faster by a factor of up to 20 on most functions, and just for the hyperbolic manifold, our distance function is a factor of $4$ faster.
For such large dimensions, RiemOpt benefits from advanced automatic transformations available in the Tensorflow backend that cannot yet be easily replicated in Julia.

\begin{figure}%
        \begin{tikzpicture}
        \begin{groupplot}[
            group style={
                group size=3 by 1,
                horizontal sep=3pt,
                y descriptions at=edge left,
            },
            axis x line*=bottom,
            axis y line*=left,
            xlabel=dimension $n$,
            ylabel={time [$\mu$s]},
            log origin=infty,
            xmode=log,ymode=log,
            xmin=2, xmax=1048576,
            width=.41\textwidth,
            ylabel shift = -1em,
            ymin=6e-4, ymax=5e3,
            ]
            \nextgroupplot[title=distance]
            \addplot[maxtime] table [x=dim, y=maxtime] \ManRnDist;
            \addplot[geomstats, autograd] table [x=dim, y=geomstats_autograd] \ManRnDist;
            \label{plots:rn:gsa}
            \addplot[geomstats, numpy] table [x=dim, y=geomstats_numpy] \ManRnDist;
            \label{plots:rn:gsn}
            \addplot[geomstats, pytorch] table [x=dim, y=geomstats_pytorch] \ManRnDist;
            \label{plots:rn:gsp}
            \addplot[geomstats, tensorflow] table [x=dim, y=geomstats_tensorflow] \ManRnDist;
            \label{plots:rn:gst}
            \addplot[geoopt] table [x=dim, y=geoopt] \ManRnDist;
            \addplot[manifolds] table [x=dim, y=manifolds] \ManRnDist;
            \label{plots:rn:mlj}
            \addplot[manopt] table [x=dim, y=manopt] \ManRnDist;
            \label{plots:rn:man}
            \addplot[pymanopt] table [x=dim, y=pymanopt] \ManRnDist;
            \label{plots:rn:pym}
            \addplot[riemopt] table [x=dim, y=riemopt] \ManRnDist;
            \label{plots:rn:rie}
            \nextgroupplot[title=exponential map]
            \addplot[maxtime] table [x=dim, y=maxtime] \ManRnExp;
            \addplot[geomstats, autograd] table [x=dim, y index=1] \ManRnExp;
            \addplot[geomstats, numpy] table [x=dim, y=geomstats_numpy] \ManRnExp;
            \addplot[geomstats, pytorch] table [x=dim, y=geomstats_pytorch] \ManRnExp;
            \addplot[geomstats, tensorflow] table [x=dim, y=geomstats_tensorflow] \ManRnExp;
            \addplot[geoopt] table [x=dim, y=geoopt] \ManRnExp;
            \label{plots:rn:geo}
            \addplot[manifolds] table [x=dim, y=manifolds] \ManRnExp;
            \addplot[manopt] table [x=dim, y=manopt] \ManRnExp;
            \addplot[pymanopt] table [x=dim, y=pymanopt] \ManRnExp;
            \addplot[riemopt] table [x=dim, y=riemopt] \ManRnExp;
            \addplot[roptlib] table [x=dim, y=roptlib] \ManRnExp;
            \label{plots:rn:rop}
            \nextgroupplot[title=logarithmic map]
            \addplot[maxtime] table [x=dim, y=maxtime] \ManRnLog;
            \addplot[geomstats, autograd] table [x=dim, y index=1] \ManRnLog;
            \addplot[geomstats, numpy] table [x=dim, y=geomstats_numpy] \ManRnLog;
            \addplot[geomstats, pytorch] table [x=dim, y=geomstats_pytorch] \ManRnLog;
            \addplot[geomstats, tensorflow] table [x=dim, y=geomstats_tensorflow] \ManRnLog;
            \addplot[geoopt] table [x=dim, y=geoopt] \ManRnLog;
            \addplot[manifolds] table [x=dim, y=manifolds] \ManRnLog;
            \addplot[manopt] table [x=dim, y=manopt] \ManRnLog;
            \addplot[pymanopt] table [x=dim, y=pymanopt] \ManRnLog;
            \addplot[riemopt] table [x=dim, y=riemopt] \ManRnLog;
        \end{groupplot}
        \matrix[
            matrix of nodes,
            anchor=south,
            draw=none,
            inner sep=0.2em,
            nodes={align=left, text width=1.25cm},
        ] at (current bounding box.north) {
        &&\ref{plots:rn:gsa}& {\hspace*{-.6cm}\small Geomstats {\\[-.6\baselineskip]\tiny (Autograd)}}& [5pt]
        \ref{plots:rn:gsn}& {\hspace*{-.6cm}\small Geomstats {\\[-.6\baselineskip]\tiny (NumPy)}}&[5pt]
        \ref{plots:rn:gsp}& {\hspace*{-.6cm}\small Geomstats {\\[-.6\baselineskip]\tiny (PyTorch)}}&[5pt]
        \ref{plots:rn:gst}& {\hspace*{-.6cm}\small Geomstats {\\[-.6\baselineskip]\tiny (Tensorflow)}}&[5pt]
        \ref{plots:rn:geo}& {\hspace*{-.6cm}\small Geoopt}&[5pt]\\
        &&\ref{plots:rn:mlj}& {\hspace*{-.6cm}\small Manifolds.jl} & [5pt]
        \ref{plots:rn:man}& {\hspace*{-.6cm}\small Manopt}&[5pt]
        \ref{plots:rn:pym}& {\hspace*{-.6cm}\small pymanopt}&[5pt]
        \ref{plots:rn:rie}& {\hspace*{-.6cm}\small Riemopt}&[5pt]
        \ref{plots:rn:rop}& {\hspace*{-.6cm}\small ROPTLIB}&[5pt]\\
        &&&&&&&\hspace{-.7cm}Benchmark &\hspace*{-.6cm}on $\mathbb R^n$\\
        };
    \end{tikzpicture}
        \begin{tikzpicture}
        \begin{groupplot}[
            group style={
                group size=3 by 1,
                horizontal sep=3pt,
                y descriptions at=edge left,
            },
            axis x line*=bottom,
            axis y line*=left,
            log origin=infty,
            xlabel=dimension $n$,
            ylabel={time [$\mu$s]},
            xmode=log,ymode=log,
            xmin=2, xmax=1048576,
            width=.41\textwidth,
            ylabel shift = -1em,
            ymin=8e-3, ymax=1e7,
        ]
            \nextgroupplot[title=distance]
            \addplot[maxtime] table [x=dim, y=maxtime] \ManHnDist;
            \addplot[geomstats, autograd] table [x index=0, y index=1] \ManHnDist;
            \addplot[geomstats, numpy] table [x index=0, y=geomstats_numpy] \ManHnDist;
            \addplot[geomstats, pytorch] table [x index=0, y=geomstats_pytorch] \ManHnDist;
            \addplot[geomstats, tensorflow] table [x index=0, y=geomstats_tensorflow] \ManHnDist;
            \addplot[geoopt] table [x index=0, y=geoopt] \ManHnDist;
            \addplot[manifolds] table [x index=0, y=manifolds] \ManHnDist;
            \addplot[manopt] table [x index=0, y=manopt] \ManHnDist;
            \addplot[pymanopt] table [x index=0, y=pymanopt] \ManHnDist;
            \addplot[riemopt] table [x index=0, y=riemopt] \ManHnDist;
            \nextgroupplot[title=exponential map]
            \addplot[maxtime] table [x=dim, y=maxtime] \ManHnExp;
            \addplot[geomstats, autograd] table [x index=0, y index=1] \ManHnExp;
            \addplot[geomstats, numpy] table [x index=0, y=geomstats_numpy] \ManHnExp;
            \addplot[geomstats, pytorch] table [x index=0, y=geomstats_pytorch] \ManHnExp;
            \addplot[geomstats, tensorflow] table [x index=0, y=geomstats_tensorflow] \ManHnExp;
            \addplot[geoopt] table [x index=0, y=geoopt] \ManHnExp;
            \addplot[manifolds] table [x index=0, y=manifolds] \ManHnExp;
            \addplot[manopt] table [x index=0, y=manopt] \ManHnExp;
            \addplot[pymanopt] table [x index=0, y=pymanopt] \ManHnExp;
            \addplot[riemopt] table [x index=0, y=riemopt] \ManHnExp;
            \addplot[roptlib] table [x index=0, y=roptlib] \ManHnExp;
            \nextgroupplot[title=logarithmic map]
            \addplot[maxtime] table [x=dim, y=maxtime] \ManHnLog;
            \addplot[geomstats, autograd] table [x index=0, y index=1] \ManHnLog;
            \addplot[geomstats, numpy] table [x index=0, y=geomstats_numpy] \ManHnLog;
            \addplot[geomstats, pytorch] table [x index=0, y=geomstats_pytorch] \ManHnLog;
            \addplot[geomstats, tensorflow] table [x index=0, y=geomstats_tensorflow] \ManHnLog;
            \addplot[geoopt] table [x index=0, y=geoopt] \ManHnLog;
            \addplot[manifolds] table [x index=0, y=manifolds] \ManHnLog;
            \addplot[manopt] table [x index=0, y=manopt] \ManHnLog;
            \addplot[pymanopt] table [x index=0, y=pymanopt] \ManHnLog;
            \addplot[riemopt] table [x index=0, y=riemopt] \ManHnLog;
        \end{groupplot}
        \matrix[
            matrix of nodes,
            anchor=south,
            draw=none,
            inner sep=0.2em,
        ] at (current bounding box.north) {
        {Benchmarks on $\mathbb H^n$}
        \\
        };
    \end{tikzpicture}
        \begin{tikzpicture}
        \begin{groupplot}[
            group style={
                group size=3 by 1,
                horizontal sep=3pt,
                y descriptions at=edge left,
            },
            axis x line*=bottom,
            axis y line*=left,
            log origin=infty,
            xlabel=dimension $n$,
            ylabel={time [$\mu$s]},
            xmode=log,ymode=log,
            xmin=2, xmax=1048576,
            width=.41\textwidth,
            ylabel shift = -1em,
            ymin=2e-3, ymax=1.8e5,
        ]
            \nextgroupplot[title=distance]
            \addplot[maxtime] table [x=dim, y=maxtime] \ManSnDist;
            \addplot[geomstats, autograd] table [x index=0, y index=1] \ManSnDist;
            \addplot[geomstats, numpy] table [x index=0, y=geomstats_numpy] \ManSnDist;
            \addplot[geomstats, pytorch] table [x index=0, y=geomstats_pytorch] \ManSnDist;
            \addplot[geomstats, tensorflow] table [x index=0, y=geomstats_tensorflow] \ManSnDist;
            \addplot[geoopt] table [x index=0, y=geoopt] \ManSnDist;
            \addplot[manifolds] table [x index=0, y=manifolds] \ManSnDist;
            \addplot[manopt] table [x index=0, y=manopt] \ManSnDist;
            \addplot[pymanopt] table [x index=0, y=pymanopt] \ManSnDist;
            \addplot[riemopt] table [x index=0, y=riemopt] \ManSnDist;
            \nextgroupplot[title=exponential map]
            \addplot[maxtime] table [x=dim, y=maxtime] \ManSnExp;
            \addplot[geomstats, autograd] table [x index=0, y index=1] \ManSnExp;
            \addplot[geomstats, numpy] table [x index=0, y=geomstats_numpy] \ManSnExp;
            \addplot[geomstats, pytorch] table [x index=0, y=geomstats_pytorch] \ManSnExp;
            \addplot[geomstats, tensorflow] table [x index=0, y=geomstats_tensorflow] \ManSnExp;
            \addplot[geoopt] table [x index=0, y=geoopt] \ManSnExp;
            \addplot[manifolds] table [x index=0, y=manifolds] \ManSnExp;
            \addplot[manopt] table [x index=0, y=manopt] \ManSnExp;
            \addplot[pymanopt] table [x index=0, y=pymanopt] \ManSnExp;
            \addplot[riemopt] table [x index=0, y=riemopt] \ManSnExp;
            \addplot[roptlib] table [x index=0, y=roptlib] \ManSnExp;
            \nextgroupplot[title=logarithmic map]
            \addplot[maxtime] table [x=dim, y=maxtime] \ManSnLog;
            \addplot[geomstats, autograd] table [x index=0, y index=1] \ManSnLog;
            \addplot[geomstats, numpy] table [x index=0, y=geomstats_numpy] \ManSnLog;
            \addplot[geomstats, pytorch] table [x index=0, y=geomstats_pytorch] \ManSnLog;
            \addplot[geomstats, tensorflow] table [x index=0, y=geomstats_tensorflow] \ManSnLog;
            \addplot[geoopt] table [x index=0, y=geoopt] \ManSnLog;
            \addplot[manifolds] table [x index=0, y=manifolds] \ManSnLog;
            \addplot[manopt] table [x index=0, y=manopt] \ManSnLog;
            \addplot[pymanopt] table [x index=0, y=pymanopt] \ManSnLog;
            \addplot[riemopt] table [x index=0, y=riemopt] \ManSnLog;
        \end{groupplot}
        \matrix[
            matrix of nodes,
            anchor=south,
            draw=none,
            inner sep=0.2em,
        ] at (current bounding box.north) {
        {Benchmarks on $\mathbb S^n$}
        \\
        };
    \end{tikzpicture}
    \caption{Benchmark of distance (left), exponential maps (middle), and logarithmic maps (right) on constant curvature manifolds $\Rn{n}$ (top) $\Hn{n}$ (middle), and $\Sn{n}$ for different dimensions $n$.
    The different colors correspond to the different software packages, where for Geomstats additionally the different line styles refer to different backends.}
    \label{fig:benchmark}
\end{figure}

For the second experiment, the results are shown in shown in Figure~\ref{fig:benchmark2}.
Compared to ROPTLIB, we see the same effect: often the Julia package is a factor of at least 2 faster.
Only for the exponential map on $\mathcal P(n)$ ROPTLIB is a factor of 2 faster, probably again due to different linear algebra library or memory management.

\begin{figure}%
    \normalsize
        \begin{tikzpicture}
        \begin{groupplot}[
            group style={
                group size=3 by 1,
                horizontal sep=3pt,
                y descriptions at=edge left,
            },
            axis x line*=bottom,
            axis y line*=left,
            log origin=infty,
            log origin=infty,
            xlabel=dimension $n$,
            ylabel={time [$\mu$s]},
            xmode=log,ymode=log,
            xtick = {2,3,4,8,16,32},
            xticklabels = {$2$, $3$, $4$, $8$, $16$,$32$},
            xmin=1.9, xmax=35,
            width=.41\textwidth,
            ylabel shift = -1em,
            ymin=8e-3,ymax=1e5,
        ]
            \nextgroupplot[title=distance]
            \addplot[maxtime] table [x=dim, y=maxtime] \ManSODist;
            \addplot[geomstats, autograd] table [x index=0, y index=1] \ManSODist;
            \addplot[geomstats, numpy] table [x index=0, y=geomstats_numpy] \ManSODist;
            \addplot[geomstats, pytorch] table [x index=0, y=geomstats_pytorch] \ManSODist;
            \addplot[geomstats, tensorflow] table [x index=0, y=geomstats_tensorflow] \ManSODist;
            \addplot[geoopt] table [x index=0, y=geoopt] \ManSODist;
            \addplot[manifolds] table [x index=0, y=manifolds] \ManSODist;
            \label{plots:son:mlj}
            \addplot[manopt] table [x index=0, y=manopt] \ManSODist;
            \label{plots:son:man}
            \addplot[pymanopt] table [x index=0, y=pymanopt] \ManSODist;
            \label{plots:son:pym}
            \addplot[riemopt] table [x index=0, y=riemopt] \ManSODist;
            \label{plots:son:rie}
            \addplot[roptlib] table [x index=0, y=roptlib] \ManSODist;
            \nextgroupplot[title=exponential map]
            \addplot[maxtime] table [x=dim, y=maxtime] \ManSOExp;
            \addplot[geomstats, autograd] table [x index=0, y index=1] \ManSOExp;
            \label{plots:son:gsa}
            \addplot[geomstats, numpy] table [x index=0, y=geomstats_numpy] \ManSOExp;
            \label{plots:son:gsn}
            \addplot[geomstats, pytorch] table [x index=0, y=geomstats_pytorch] \ManSOExp;
            \label{plots:son:gsp}
            \addplot[geomstats, tensorflow] table [x index=0, y=geomstats_tensorflow] \ManSOExp;
            \label{plots:son:gst}
            \addplot[geoopt] table [x index=0, y=geoopt] \ManSOExp;
            \addplot[manifolds] table [x index=0, y=manifolds] \ManSOExp;
            \addplot[manopt] table [x index=0, y=manopt] \ManSOExp;
            \addplot[pymanopt] table [x index=0, y=pymanopt] \ManSOExp;
            \addplot[riemopt] table [x index=0, y=riemopt] \ManSOExp;
            \addplot[roptlib] table [x index=0, y=roptlib] \ManSOExp;
            \label{plots:son:rop}
            \nextgroupplot[title=logarithmic map]
            \addplot[maxtime] table [x=dim, y=maxtime] \ManSOLog;
            \addplot[geomstats, autograd] table [x index=0, y index=1] \ManSOLog;
            \addplot[geomstats, numpy] table [x index=0, y=geomstats_numpy] \ManSOLog;
            \addplot[geomstats, pytorch] table [x index=0, y=geomstats_pytorch] \ManSOLog;
            \addplot[geomstats, tensorflow] table [x index=0, y=geomstats_tensorflow] \ManSOLog;
            \addplot[geoopt] table [x index=0, y=geoopt] \ManSOLog;
            \addplot[manifolds] table [x index=0, y=manifolds] \ManSOLog;
            \addplot[manopt] table [x index=0, y=manopt] \ManSOLog;
            \addplot[pymanopt] table [x index=0, y=pymanopt] \ManSOLog;
            \addplot[riemopt] table [x index=0, y=riemopt] \ManSOLog;
            \addplot[roptlib] table [x index=0, y=roptlib] \ManSOLog;
        \end{groupplot}
        \matrix[
            matrix of nodes,
            anchor=south,
            draw=none,
            inner sep=0.2em,
            nodes={align=left, text width=1.25cm},
        ] at (current bounding box.north) {
        &&\ref{plots:son:gsa}& {\hspace*{-.6cm}\small Geomstats {\\[-.6\baselineskip]\tiny (Autograd)}}& [5pt]
        \ref{plots:son:gsn}& {\hspace*{-.6cm}\small Geomstats {\\[-.6\baselineskip]\tiny (NumPy)}}&[5pt]
        \ref{plots:son:gsp}& {\hspace*{-.6cm}\small Geomstats {\\[-.6\baselineskip]\tiny (PyTorch)}}&[5pt]
        \ref{plots:son:gst}& {\hspace*{-.6cm}\small Geomstats {\\[-.6\baselineskip]\tiny (Tensorflow)}}&[5pt]
        \ref{plots:rn:geo}& {\hspace*{-.6cm}\small Geoopt}&[5pt]
        \\
        &&\ref{plots:son:mlj}& {\hspace*{-.6cm}\small Manifolds.jl} & [5pt]
        \ref{plots:son:man}& {\hspace*{-.6cm}\small Manopt}&[5pt]
        \ref{plots:son:pym}& {\hspace*{-.6cm}\small pymanopt}&[5pt]
        \ref{plots:son:rie}& {\hspace*{-.6cm}\small Riemopt}&[5pt]
        \ref{plots:rn:rop}& {\hspace*{-.6cm}\small ROPTLIB}&[5pt]\\
        &&&&&&&\hspace{-.7cm}Benchmark &\hspace*{-.6cm}on $\mathrm{SO}(n)$\\
        };
    \end{tikzpicture}
        \begin{tikzpicture}
        \begin{groupplot}[
            group style={
                group size=3 by 1,
                horizontal sep=3pt,
                y descriptions at=edge left,
            },
            axis x line*=bottom,
            axis y line*=left,
            xlabel=dimension $n$,
            ylabel={time [$\mu$s]},
            log origin=infty,
            xmode=log,ymode=log,
            xtick = {2,3,4,8,16,32},
            xticklabels = {$2$, $3$, $4$, $8$, $16$,$32$},
            xmin=1.9, xmax=35,
            ymin=1e-2,ymax=2e5,
            width=.41\textwidth,
            ylabel shift = -1em,
        ]
            \nextgroupplot[title=distance]
            \addplot[maxtime] table [x=dim, y=maxtime] \ManSPDDist;
            \addplot[geomstats, autograd] table [x index=0, y index=1] \ManSPDDist;
            \addplot[geomstats, numpy] table [x index=0, y=geomstats_numpy] \ManSPDDist;
            \addplot[geomstats, pytorch] table [x index=0, y=geomstats_pytorch] \ManSPDDist;
            \addplot[geomstats, tensorflow] table [x index=0, y=geomstats_tensorflow] \ManSPDDist;
            \addplot[geoopt] table [x index=0, y=geoopt] \ManSPDDist;
            \addplot[manifolds] table [x index=0, y=manifolds] \ManSPDDist;
            \addplot[manopt] table [x index=0, y=manopt] \ManSPDDist;
            \addplot[pymanopt] table [x index=0, y=pymanopt] \ManSPDDist;
            \addplot[riemopt] table [x index=0, y=riemopt] \ManSPDDist;
            \nextgroupplot[title=exponential map]
            \addplot[maxtime] table [x=dim, y=maxtime] \ManSPDExp;
            \addplot[geomstats, autograd] table [x index=0, y index=1] \ManSPDExp;
            \addplot[geomstats, numpy] table [x index=0, y=geomstats_numpy] \ManSPDExp;
            \addplot[geomstats, pytorch] table [x index=0, y=geomstats_pytorch] \ManSPDExp;
            \addplot[geomstats, tensorflow] table [x index=0, y=geomstats_tensorflow] \ManSPDExp;
            \addplot[geoopt] table [x index=0, y=geoopt] \ManSPDExp;
            \addplot[manifolds] table [x index=0, y=manifolds] \ManSPDExp;
            \addplot[manopt] table [x index=0, y=manopt] \ManSPDExp;
            \addplot[pymanopt] table [x index=0, y=pymanopt] \ManSPDExp;
            \addplot[riemopt] table [x index=0, y=riemopt] \ManSPDExp;
            \addplot[roptlib] table [x index=0, y=roptlib] \ManSPDExp;
            \nextgroupplot[title=logarithmic map]
            \addplot[maxtime] table [x=dim, y=maxtime] \ManSPDLog;
            \addplot[geomstats, autograd] table [x index=0, y index=1] \ManSPDLog;
            \addplot[geomstats, numpy] table [x index=0, y=geomstats_numpy] \ManSPDLog;
            \addplot[geomstats, pytorch] table [x index=0, y=geomstats_pytorch] \ManSPDLog;
            \addplot[geomstats, tensorflow] table [x index=0, y=geomstats_tensorflow] \ManSPDLog;
            \addplot[geoopt] table [x index=0, y=geoopt] \ManSPDLog;
            \addplot[manifolds] table [x index=0, y=manifolds] \ManSPDLog;
            \addplot[manopt] table [x index=0, y=manopt] \ManSPDLog;
            \addplot[pymanopt] table [x index=0, y=pymanopt] \ManSPDLog;
            \addplot[riemopt] table [x index=0, y=riemopt] \ManSPDLog;
        \end{groupplot}
        \matrix[
            matrix of nodes,
            anchor=south,
            draw=none,
            inner sep=0.2em,
        ] at (current bounding box.north) {
        {Benchmarks on $\mathcal P(n)$} \\
        };
    \end{tikzpicture}
        \begin{tikzpicture}
        \begin{groupplot}[
            group style={
                group size=3 by 1,
                horizontal sep=3pt,
                y descriptions at=edge left,
            },
            axis x line*=bottom,
            axis y line*=left,
            xlabel=dimension $n$,
            ylabel={time [$\mu$s]},
            log origin=infty,
            xmode=log,ymode=log,
            xtick = {2,3,4,8,16,32},
            xticklabels = {$2$, $3$, $4$, $8$, $16$,$32$},
            xmin=1.9, xmax=35,
            width=.41\textwidth,
            ylabel shift = -.7em,
            ymin=1e2, ymax=2e8
        ]
            \nextgroupplot[title=distance]
            \addplot[maxtime] table [x=dim, y=maxtime] \ManPowDist;
            \addplot[geomstats, autograd] table [x index=0, y index=1] \ManPowDist;
            \addplot[geomstats, numpy] table [x index=0, y=geomstats_numpy] \ManPowDist;
            \addplot[geomstats, pytorch] table [x index=0, y=geomstats_pytorch] \ManPowDist;
            \addplot[geomstats, tensorflow] table [x index=0, y=geomstats_tensorflow] \ManPowDist;
            \addplot[geoopt] table [x index=0, y=geoopt] \ManPowDist;
            \addplot[manifolds] table [x index=0, y=manifolds] \ManPowDist;
            \addplot[manopt] table [x index=0, y=manopt] \ManPowDist;
            \addplot[pymanopt] table [x index=0, y=pymanopt] \ManPowDist;
            \addplot[riemopt] table [x index=0, y=riemopt] \ManPowDist;
            \nextgroupplot[title=exponential map]
            \addplot[maxtime] table [x=dim, y=maxtime] \ManPowExp;
            \addplot[geomstats, autograd] table [x index=0, y index=1] \ManPowExp;
            \addplot[geomstats, numpy] table [x index=0, y=geomstats_numpy] \ManPowExp;
            \addplot[geomstats, pytorch] table [x index=0, y=geomstats_pytorch] \ManPowExp;
            \addplot[geomstats, tensorflow] table [x index=0, y=geomstats_tensorflow] \ManPowExp;
            \addplot[geoopt] table [x index=0, y=geoopt] \ManPowExp;
            \addplot[manifolds] table [x index=0, y=manifolds] \ManPowExp;
            \addplot[manopt] table [x index=0, y=manopt] \ManPowExp;
            \addplot[pymanopt] table [x index=0, y=pymanopt] \ManPowExp;
            \addplot[riemopt] table [x index=0, y=riemopt] \ManPowExp;
            \addplot[roptlib] table [x index=0, y=roptlib] \ManPowExp;
            \nextgroupplot[title=logarithmic map]
            \addplot[maxtime] table [x=dim, y=maxtime] \ManPowLog;
            \addplot[geomstats, autograd] table [x index=0, y index=1] \ManPowLog;
            \addplot[geomstats, numpy] table [x index=0, y=geomstats_numpy] \ManPowLog;
            \addplot[geomstats, pytorch] table [x index=0, y=geomstats_pytorch] \ManPowLog;
            \addplot[geomstats, tensorflow] table [x index=0, y=geomstats_tensorflow] \ManPowLog;
            \addplot[geoopt] table [x index=0, y=geoopt] \ManPowLog;
            \addplot[manifolds] table [x index=0, y=manifolds] \ManPowLog;
            \addplot[manopt] table [x index=0, y=manopt] \ManPowLog;
            \addplot[pymanopt] table [x index=0, y=pymanopt] \ManPowLog;
            \addplot[riemopt] table [x index=0, y=riemopt] \ManPowLog;
        \end{groupplot}
        \matrix[
            matrix of nodes,
            anchor=south,
            draw=none,
            inner sep=0.2em,
        ] at (current bounding box.north) {
        {Benchmarks on $(\mathcal P(n))^{128\times 128}$} \\
        };
    \end{tikzpicture}
    \caption{Benchmark of distance (left), exponential maps (middle), and logarithmic maps (right) on the manifolds of rotations (top), manifold of symmetric positive definite matrices (middle), and its power manifold (bottom).
    The different colors correspond to the different software packages, where for Geomstats additionally the different line styles refer to different backends.}
    \label{fig:benchmark2}
\end{figure}

Concerning RiemOpt, again, only for the very high-dimensional power manifold, this Tensorflow based package is usually a factor of approximately 10 faster.
This starts already for the case $n=3$, where \lstinline!Manifolds.jl! is a factor of $2$ slower for exponential and logarithmic maps but a factor of 5 faster for distance; all other packages are at least a factor of 4-5 slower, while Manopt seems to be the slowest by several orders of magnitude, probably due to the use of cell arrays in Matlab.

For the highest dimension $n=32$ and the power manifold, both Pymanopt and Manopt reach the same order of magnitude as \lstinline!Manifolds.jl! and Riemopt, where the latter is still a factor of 5 faster.

\lstinline!Manifolds.jl! is the fastest package overall -- on par with ROPTLIB wherever it provides an implementation -- and only significantly surpassed by Riemopt for some large-scale manifolds of dimensions of more than $2^{16}$.

We further compared the accuracy of the methods run in the benchmark.
We first computed reference results using \lstinline!Manifolds.jl!
with \lstinline!BigFloat! as number type and both \lstinline!GenericLinearAlgebra.jl! and \lstinline!GenericSchur.jl! for the more precise linear algebra routines.
We then performed the accuracy test against these exact results for all above benchmarked functions and packages. The full results are given in Appendix~\ref{app:Accuracy}.

Overall, \lstinline!Manifolds.jl! lies among the best performing ones here as well. Only for the symmetric positive definite matrices $\mathcal P(n)$ -- more precisely the logarithmic and exponential map thereon -- our library is slightly less accurate than others. The PyTorch backend seems to only reach a significantly lower accuracy, apparently due to an internal conversion to 32-bit floating point representation.
\section{Conclusions}
\label{sec:conclusions}

\lstinline!Manifolds.jl! provides a comprehensive API and set of features for working with points on manifolds.
It also offers one of the largest collections of implementations of manifolds.
Using this API, \lstinline!Manopt.jl! provides a library of algorithms for optimization on manifolds, and \lstinline!ManifoldDiffEq.jl! offers a variety of solvers for ordinary differential equations.
As a result, all manifolds defined with \lstinline!ManifoldsBase.jl! are compatible with both libraries and can directly be used therein.

One prominent advantage of \lstinline!Manifolds.jl! over similar packages is that, by using Julia, the “two-language-barrier” is overcome while keeping the code concise and expressive.
Implementations in \lstinline!Manifolds.jl! were demonstrated to be superior in performance to those in libraries with interfaces in Python and Matlab, where only for large scale manifolds Riemopt outperforms \lstinline!Manifolds.jl! slightly.
The library is approximately twice as fast as ROPTLIB, though that is probably due to the linear algebra library.

However, this high performance did not require sacrificing clarity of the implementations themselves, the ease of use of a dynamic programming language, or computational accuracy.

Future work on \lstinline!Manifolds.jl! will improve integration of Julia's automatic differentiation packages, add support for GPU acceleration, and expand the set of available manifolds and operations.
Other plans include support for more statistical and machine learning methods.

\begin{acks}
  We would like to thank the anonymous reviewers for their valuable feedback.
  \\
  SDA was supported by the National Science Foundation under grant number 1650113 and the Deutsche Forschungsgemeinschaft (DFG, German Research Foundation) under Germany’s Excellence Strategy – EXC number 2064/1 – Project number 390727645.
  MB and KR were supported by Foundation for Polish Science co-financed by the European Union under the European Regional Development Fund within TEAM-NET programme under grant POIR.04.04.00-00-15E5/18.
\end{acks}

\bibliography{bibliography}

\newpage

\appendix

\section{Accuracy Comparisons}\label{app:Accuracy}

Accuracy of computations performed by different libraries is compared in Fig.~\ref{fig:accuracy} and Fig.~\ref{fig:accuracy2}.
In all cases 10 points and tangent vectors were randomly sampled from those used for benchmarking.
Reference results were obtained in \lstinline!Manifolds.jl! using computations on arbitrary precision numbers (\lstinline!BigFloat!) instead of \lstinline!Float64!.
Packages \lstinline!GenericLinearAlgebra.jl! and \lstinline!GenericSchur.jl! were used to supply linear algebra routines compatible with \lstinline!BigFloat!. Additionally, generic matrix exponential and logarithm were implemented using generic eigen decomposition.

On $\Rn{n}$ all libraries exhibit similar accuracy, except for distance calculation on high-dimensional manifolds. There is, however, little correlation between accuracy and performance.

On $\Hn{n}$ geomstats is the least accurate and the slowest. Geoopt appears to return wrong results of exponential map. Manopt and \lstinline!Manifolds.jl! implement a more accurate logarithmic map than other libraries, which is still fast to compute.

On $\Sn{n}$ geomstats appears to convert intermediate results to a 32-bit floating point type with PyTorch and Tensorflow backends. Apparent high accuracy of exponential map with autograd backend in the high-dimensional case is likely an artifact caused by estimating accuracy based on 10 points. Logarithmic map in geoopt was less accurate than in other libraries.

Accuracy of $\SOn{n}$ was roughly similar across all libraries except for exponential map on $\SOn{2}$ in geomstats with Numpy and logarithmic map with Tensorflow.

On $\symposdef{n}$ geomstats with Pytorch is the least accurate. \lstinline!Manifolds.jl! has in some cases less accurate exponential and logarithmic maps than other libraries. These results roughly translate to the power manifold $\symposdef{n}^{128\times 128}$. Accuracy on $\symposdef{32}^{128\times 128}$ is not included because calculating reference results would require large computational resources, while these results would be unlikely to provide any new insight.

\begin{figure}%
        \begin{tikzpicture}
        \begin{groupplot}[
            group style={
                group size=3 by 1,
                horizontal sep=3pt,
                y descriptions at=edge left,
            },
            axis x line*=bottom,
            axis y line*=left,
            xlabel=dimension $n$,
            ylabel={mean absolute error},
            log origin=infty,
            xmode=log,ymode=log,
            xmin=2, xmax=1048576,
            width=.41\textwidth,
            ylabel shift = -1em,
            ymin=1e-17, ymax=1e-10,
            ]
            \nextgroupplot[title=distance]
            \addplot[maxacc] table [x=dim, y=maxacc] \ManRnDistAcc;
            \addplot[geomstats, autograd] table [x=dim, y=geomstats_autograd] \ManRnDistAcc;
            \label{plots:rn:gsa}
            \addplot[geomstats, numpy] table [x=dim, y=geomstats_numpy] \ManRnDistAcc;
            \label{plots:rn:gsn}
            \addplot[geomstats, pytorch] table [x=dim, y=geomstats_pytorch] \ManRnDistAcc;
            \label{plots:rn:gsp}
            \addplot[geomstats, tensorflow] table [x=dim, y=geomstats_tensorflow] \ManRnDistAcc;
            \label{plots:rn:gst}
            \addplot[geoopt] table [x=dim, y=geoopt] \ManRnDistAcc;
            \addplot[manifolds] table [x=dim, y=manifolds] \ManRnDistAcc;
            \label{plots:rn:mlj}
            \addplot[manopt] table [x=dim, y=manopt] \ManRnDistAcc;
            \label{plots:rn:man}
            \addplot[pymanopt] table [x=dim, y=pymanopt] \ManRnDistAcc;
            \label{plots:rn:pym}
            \addplot[riemopt] table [x=dim, y=riemopt] \ManRnDistAcc;
            \label{plots:rn:rie}
            \nextgroupplot[title=exponential map]
            \addplot[maxacc] table [x=dim, y=maxacc] \ManRnExpAcc;
            \addplot[geomstats, autograd] table [x=dim, y index=1] \ManRnExpAcc;
            \addplot[geomstats, numpy] table [x=dim, y=geomstats_numpy] \ManRnExpAcc;
            \addplot[geomstats, pytorch] table [x=dim, y=geomstats_pytorch] \ManRnExpAcc;
            \addplot[geomstats, tensorflow] table [x=dim, y=geomstats_tensorflow] \ManRnExpAcc;
            \addplot[geoopt] table [x=dim, y=geoopt] \ManRnExpAcc;
            \label{plots:rn:geo}
            \addplot[manifolds] table [x=dim, y=manifolds] \ManRnExpAcc;
            \addplot[manopt] table [x=dim, y=manopt] \ManRnExpAcc;
            \addplot[pymanopt] table [x=dim, y=pymanopt] \ManRnExpAcc;
            \addplot[riemopt] table [x=dim, y=riemopt] \ManRnExpAcc;
            \addplot[roptlib] table [x=dim, y=roptlib] \ManRnExpAcc;
            \label{plots:rn:rop}
            \nextgroupplot[title=logarithmic map]
            \addplot[maxacc] table [x=dim, y=maxacc] \ManRnLogAcc;
            \addplot[geomstats, autograd] table [x=dim, y index=1] \ManRnLogAcc;
            \addplot[geomstats, numpy] table [x=dim, y=geomstats_numpy] \ManRnLogAcc;
            \addplot[geomstats, pytorch] table [x=dim, y=geomstats_pytorch] \ManRnLogAcc;
            \addplot[geomstats, tensorflow] table [x=dim, y=geomstats_tensorflow] \ManRnLogAcc;
            \addplot[geoopt] table [x=dim, y=geoopt] \ManRnLogAcc;
            \addplot[manifolds] table [x=dim, y=manifolds] \ManRnLogAcc;
            \addplot[manopt] table [x=dim, y=manopt] \ManRnLogAcc;
            \addplot[pymanopt] table [x=dim, y=pymanopt] \ManRnLogAcc;
            \addplot[riemopt] table [x=dim, y=riemopt] \ManRnLogAcc;
        \end{groupplot}
        \matrix[
            matrix of nodes,
            anchor=south,
            draw=none,
            inner sep=0.2em,
            nodes={align=left, text width=1.25cm},
        ] at (current bounding box.north) {
        &&\ref{plots:rn:gsa}& {\hspace*{-.6cm}\small Geomstats {\\[-.6\baselineskip]\tiny (Autograd)}}& [5pt]
        \ref{plots:rn:gsn}& {\hspace*{-.6cm}\small Geomstats {\\[-.6\baselineskip]\tiny (NumPy)}}&[5pt]
        \ref{plots:rn:gsp}& {\hspace*{-.6cm}\small Geomstats {\\[-.6\baselineskip]\tiny (PyTorch)}}&[5pt]
        \ref{plots:rn:gst}& {\hspace*{-.6cm}\small Geomstats {\\[-.6\baselineskip]\tiny (Tensorflow)}}&[5pt]
        \ref{plots:rn:geo}& {\hspace*{-.6cm}\small Geoopt}&[5pt]\\
        &&\ref{plots:rn:mlj}& {\hspace*{-.6cm}\small Manifolds.jl} & [5pt]
        \ref{plots:rn:man}& {\hspace*{-.6cm}\small Manopt}&[5pt]
        \ref{plots:rn:pym}& {\hspace*{-.6cm}\small pymanopt}&[5pt]
        \ref{plots:rn:rie}& {\hspace*{-.6cm}\small Riemopt}&[5pt]
        \ref{plots:rn:rop}& {\hspace*{-.6cm}\small ROPTLIB}&[5pt]\\
        &&&&&&\hspace{-.3cm}Accuracy &\hspace*{-.3cm}on $\mathbb R^n$\\
        };
    \end{tikzpicture}
        \begin{tikzpicture}
        \begin{groupplot}[
            group style={
                group size=3 by 1,
                horizontal sep=3pt,
                y descriptions at=edge left,
            },
            axis x line*=bottom,
            axis y line*=left,
            log origin=infty,
            xlabel=dimension $n$,
            ylabel={mean absolute error},
            xmode=log,ymode=log,
            xmin=2, xmax=1048576,
            width=.41\textwidth,
            ylabel shift = -1em,
            ymin=1e-17, ymax=5e3,
        ]
            \nextgroupplot[title=distance]
            \addplot[maxacc] table [x=dim, y=maxacc] \ManHnDistAcc;
            \addplot[geomstats, autograd] table [x index=0, y index=1] \ManHnDistAcc;
            \addplot[geomstats, numpy] table [x index=0, y=geomstats_numpy] \ManHnDistAcc;
            \addplot[geomstats, pytorch] table [x index=0, y=geomstats_pytorch] \ManHnDistAcc;
            \addplot[geomstats, tensorflow] table [x index=0, y=geomstats_tensorflow] \ManHnDistAcc;
            \addplot[geoopt] table [x index=0, y=geoopt] \ManHnDistAcc;
            \addplot[manifolds] table [x index=0, y=manifolds] \ManHnDistAcc;
            \addplot[manopt] table [x index=0, y=manopt] \ManHnDistAcc;
            \addplot[pymanopt] table [x index=0, y=pymanopt] \ManHnDistAcc;
            \addplot[riemopt] table [x index=0, y=riemopt] \ManHnDistAcc;
            \nextgroupplot[title=exponential map]
            \addplot[maxacc] table [x=dim, y=maxacc] \ManHnExpAcc;
            \addplot[geomstats, autograd] table [x index=0, y index=1] \ManHnExpAcc;
            \addplot[geomstats, numpy] table [x index=0, y=geomstats_numpy] \ManHnExpAcc;
            \addplot[geomstats, pytorch] table [x index=0, y=geomstats_pytorch] \ManHnExpAcc;
            \addplot[geomstats, tensorflow] table [x index=0, y=geomstats_tensorflow] \ManHnExpAcc;
            \addplot[geoopt] table [x index=0, y=geoopt] \ManHnExpAcc;
            \addplot[manifolds] table [x index=0, y=manifolds] \ManHnExpAcc;
            \addplot[manopt] table [x index=0, y=manopt] \ManHnExpAcc;
            \addplot[pymanopt] table [x index=0, y=pymanopt] \ManHnExpAcc;
            \addplot[riemopt] table [x index=0, y=riemopt] \ManHnExpAcc;
            \addplot[roptlib] table [x index=0, y=roptlib] \ManHnExpAcc;
            \nextgroupplot[title=logarithmic map]
            \addplot[maxacc] table [x=dim, y=maxacc] \ManHnLogAcc;
            \addplot[geomstats, autograd] table [x index=0, y index=1] \ManHnLogAcc;
            \addplot[geomstats, numpy] table [x index=0, y=geomstats_numpy] \ManHnLogAcc;
            \addplot[geomstats, pytorch] table [x index=0, y=geomstats_pytorch] \ManHnLogAcc;
            \addplot[geomstats, tensorflow] table [x index=0, y=geomstats_tensorflow] \ManHnLogAcc;
            \addplot[geoopt] table [x index=0, y=geoopt] \ManHnLogAcc;
            \addplot[manifolds] table [x index=0, y=manifolds] \ManHnLogAcc;
            \addplot[manopt] table [x index=0, y=manopt] \ManHnLogAcc;
            \addplot[pymanopt] table [x index=0, y=pymanopt] \ManHnLogAcc;
            \addplot[riemopt] table [x index=0, y=riemopt] \ManHnLogAcc;
        \end{groupplot}
        \matrix[
            matrix of nodes,
            anchor=south,
            draw=none,
            inner sep=0.2em,
        ] at (current bounding box.north) {
        {Accuracy on $\mathbb H^n$}
        \\
        };
    \end{tikzpicture}
        \begin{tikzpicture}
        \begin{groupplot}[
            group style={
                group size=3 by 1,
                horizontal sep=3pt,
                y descriptions at=edge left,
            },
            axis x line*=bottom,
            axis y line*=left,
            log origin=infty,
            xlabel=dimension $n$,
            ylabel={mean absolute error},
            xmode=log,ymode=log,
            xmin=2, xmax=1048576,
            width=.41\textwidth,
            ylabel shift = -1em,
            ymin=1e-17, ymax=5e-2,
        ]
            \nextgroupplot[title=distance]
            \addplot[maxacc] table [x=dim, y=maxacc] \ManSnDistAcc;
            \addplot[geomstats, autograd] table [x index=0, y index=1] \ManSnDistAcc;
            \addplot[geomstats, numpy] table [x index=0, y=geomstats_numpy] \ManSnDistAcc;
            \addplot[geomstats, pytorch] table [x index=0, y=geomstats_pytorch] \ManSnDistAcc;
            \addplot[geomstats, tensorflow] table [x index=0, y=geomstats_tensorflow] \ManSnDistAcc;
            \addplot[geoopt] table [x index=0, y=geoopt] \ManSnDistAcc;
            \addplot[manifolds] table [x index=0, y=manifolds] \ManSnDistAcc;
            \addplot[manopt] table [x index=0, y=manopt] \ManSnDistAcc;
            \addplot[pymanopt] table [x index=0, y=pymanopt] \ManSnDistAcc;
            \addplot[riemopt] table [x index=0, y=riemopt] \ManSnDistAcc;
            \nextgroupplot[title=exponential map]
            \addplot[maxacc] table [x=dim, y=maxacc] \ManSnExpAcc;
            \addplot[geomstats, autograd] table [x index=0, y index=1] \ManSnExpAcc;
            \addplot[geomstats, numpy] table [x index=0, y=geomstats_numpy] \ManSnExpAcc;
            \addplot[geomstats, pytorch] table [x index=0, y=geomstats_pytorch] \ManSnExpAcc;
            \addplot[geomstats, tensorflow] table [x index=0, y=geomstats_tensorflow] \ManSnExpAcc;
            \addplot[geoopt] table [x index=0, y=geoopt] \ManSnExpAcc;
            \addplot[manifolds] table [x index=0, y=manifolds] \ManSnExpAcc;
            \addplot[manopt] table [x index=0, y=manopt] \ManSnExpAcc;
            \addplot[pymanopt] table [x index=0, y=pymanopt] \ManSnExpAcc;
            \addplot[riemopt] table [x index=0, y=riemopt] \ManSnExpAcc;
            \addplot[roptlib] table [x index=0, y=roptlib] \ManSnExpAcc;
            \nextgroupplot[title=logarithmic map]
            \addplot[maxacc] table [x=dim, y=maxacc] \ManSnLogAcc;
            \addplot[geomstats, autograd] table [x index=0, y index=1] \ManSnLogAcc;
            \addplot[geomstats, numpy] table [x index=0, y=geomstats_numpy] \ManSnLogAcc;
            \addplot[geomstats, pytorch] table [x index=0, y=geomstats_pytorch] \ManSnLogAcc;
            \addplot[geomstats, tensorflow] table [x index=0, y=geomstats_tensorflow] \ManSnLogAcc;
            \addplot[geoopt] table [x index=0, y=geoopt] \ManSnLogAcc;
            \addplot[manifolds] table [x index=0, y=manifolds] \ManSnLogAcc;
            \addplot[manopt] table [x index=0, y=manopt] \ManSnLogAcc;
            \addplot[pymanopt] table [x index=0, y=pymanopt] \ManSnLogAcc;
            \addplot[riemopt] table [x index=0, y=riemopt] \ManSnLogAcc;
        \end{groupplot}
        \matrix[
            matrix of nodes,
            anchor=south,
            draw=none,
            inner sep=0.2em,
        ] at (current bounding box.north) {
        {Accuracy on $\mathbb S^n$}
        \\
        };
    \end{tikzpicture}
    \caption{Mean absolute error of distance (left), exponential maps (middle), and logarithmic maps (right) on constant curvature manifolds $\Rn{n}$ (top) $\Hn{n}$ (middle), and $\Sn{n}$ for different dimensions $n$.
    The different colors correspond to the different software packages, where for Geomstats additionally the different line styles refer to different backends.}
    \label{fig:accuracy}
\end{figure}

\begin{figure}%
    \normalsize
        \begin{tikzpicture}
        \begin{groupplot}[
            group style={
                group size=3 by 1,
                horizontal sep=3pt,
                y descriptions at=edge left,
            },
            axis x line*=bottom,
            axis y line*=left,
            log origin=infty,
            log origin=infty,
            xlabel=dimension $n$,
            ylabel={mean absolute error},
            xmode=log,ymode=log,
            xtick = {2,3,4,8,16,32},
            xticklabels = {$2$, $3$, $4$, $8$, $16$,$32$},
            xmin=1.9, xmax=35,
            width=.41\textwidth,
            ylabel shift = -1em,
            ymin=1e-17, ymax=1e-3,
        ]
            \nextgroupplot[title=distance]
            \addplot[maxacc] table [x=dim, y=maxacc] \ManSODistAcc;
            \addplot[geomstats, autograd] table [x index=0, y index=1] \ManSODistAcc;
            \addplot[geomstats, numpy] table [x index=0, y=geomstats_numpy] \ManSODistAcc;
            \addplot[geomstats, pytorch] table [x index=0, y=geomstats_pytorch] \ManSODistAcc;
            \addplot[geomstats, tensorflow] table [x index=0, y=geomstats_tensorflow] \ManSODistAcc;
            \addplot[geoopt] table [x index=0, y=geoopt] \ManSODistAcc;
            \addplot[manifolds] table [x index=0, y=manifolds] \ManSODistAcc;
            \label{plots:son:mlj}
            \addplot[manopt] table [x index=0, y=manopt] \ManSODistAcc;
            \label{plots:son:man}
            \addplot[pymanopt] table [x index=0, y=pymanopt] \ManSODistAcc;
            \label{plots:son:pym}
            \addplot[riemopt] table [x index=0, y=riemopt] \ManSODistAcc;
            \label{plots:son:rie}
            \addplot[roptlib] table [x index=0, y=roptlib] \ManSODistAcc;
            \nextgroupplot[title=exponential map]
            \addplot[maxacc] table [x=dim, y=maxacc] \ManSOExpAcc;
            \addplot[geomstats, autograd] table [x index=0, y index=1] \ManSOExpAcc;
            \label{plots:son:gsa}
            \addplot[geomstats, numpy] table [x index=0, y=geomstats_numpy] \ManSOExpAcc;
            \label{plots:son:gsn}
            \addplot[geomstats, pytorch] table [x index=0, y=geomstats_pytorch] \ManSOExpAcc;
            \label{plots:son:gsp}
            \addplot[geomstats, tensorflow] table [x index=0, y=geomstats_tensorflow] \ManSOExpAcc;
            \label{plots:son:gst}
            \addplot[geoopt] table [x index=0, y=geoopt] \ManSOExpAcc;
            \addplot[manifolds] table [x index=0, y=manifolds] \ManSOExpAcc;
            \addplot[manopt] table [x index=0, y=manopt] \ManSOExpAcc;
            \addplot[pymanopt] table [x index=0, y=pymanopt] \ManSOExpAcc;
            \addplot[riemopt] table [x index=0, y=riemopt] \ManSOExpAcc;
            \addplot[roptlib] table [x index=0, y=roptlib] \ManSOExpAcc;
            \label{plots:son:rop}
            \nextgroupplot[title=logarithmic map]
            \addplot[maxacc] table [x=dim, y=maxacc] \ManSOLogAcc;
            \addplot[geomstats, autograd] table [x index=0, y index=1] \ManSOLogAcc;
            \addplot[geomstats, numpy] table [x index=0, y=geomstats_numpy] \ManSOLogAcc;
            \addplot[geomstats, pytorch] table [x index=0, y=geomstats_pytorch] \ManSOLogAcc;
            \addplot[geomstats, tensorflow] table [x index=0, y=geomstats_tensorflow] \ManSOLogAcc;
            \addplot[geoopt] table [x index=0, y=geoopt] \ManSOLogAcc;
            \addplot[manifolds] table [x index=0, y=manifolds] \ManSOLogAcc;
            \addplot[manopt] table [x index=0, y=manopt] \ManSOLogAcc;
            \addplot[pymanopt] table [x index=0, y=pymanopt] \ManSOLogAcc;
            \addplot[riemopt] table [x index=0, y=riemopt] \ManSOLogAcc;
            \addplot[roptlib] table [x index=0, y=roptlib] \ManSOLogAcc;
        \end{groupplot}
        \matrix[
            matrix of nodes,
            anchor=south,
            draw=none,
            inner sep=0.2em,
            nodes={align=left, text width=1.25cm},
        ] at (current bounding box.north) {
        &&\ref{plots:son:gsa}& {\hspace*{-.6cm}\small Geomstats {\\[-.6\baselineskip]\tiny (Autograd)}}& [5pt]
        \ref{plots:son:gsn}& {\hspace*{-.6cm}\small Geomstats {\\[-.6\baselineskip]\tiny (NumPy)}}&[5pt]
        \ref{plots:son:gsp}& {\hspace*{-.6cm}\small Geomstats {\\[-.6\baselineskip]\tiny (PyTorch)}}&[5pt]
        \ref{plots:son:gst}& {\hspace*{-.6cm}\small Geomstats {\\[-.6\baselineskip]\tiny (Tensorflow)}}&[5pt]
        \ref{plots:rn:geo}& {\hspace*{-.6cm}\small Geoopt}&[5pt]
        \\
        &&\ref{plots:son:mlj}& {\hspace*{-.6cm}\small Manifolds.jl} & [5pt]
        \ref{plots:son:man}& {\hspace*{-.6cm}\small Manopt}&[5pt]
        \ref{plots:son:pym}& {\hspace*{-.6cm}\small pymanopt}&[5pt]
        \ref{plots:son:rie}& {\hspace*{-.6cm}\small Riemopt}&[5pt]
        \ref{plots:rn:rop}& {\hspace*{-.6cm}\small ROPTLIB}&[5pt]\\
        &&&&&&\hspace{-.6cm}Accuracy &\hspace*{-.5cm}on $\mathrm{SO}(n)$\\
        };
    \end{tikzpicture}
        \begin{tikzpicture}
        \begin{groupplot}[
            group style={
                group size=3 by 1,
                horizontal sep=3pt,
                y descriptions at=edge left,
            },
            axis x line*=bottom,
            axis y line*=left,
            xlabel=dimension $n$,
            ylabel={mean absolute error},
            log origin=infty,
            xmode=log,ymode=log,
            xtick = {2,3,4,8,16,32},
            xticklabels = {$2$, $3$, $4$, $8$, $16$,$32$},
            xmin=1.9, xmax=35,
            ymin=1e-17, ymax=5e-6,
            width=.41\textwidth,
            ylabel shift = -1em,
        ]
            \nextgroupplot[title=distance]
            \addplot[maxacc] table [x=dim, y=maxacc] \ManSPDDistAcc;
            \addplot[geomstats, autograd] table [x index=0, y index=1] \ManSPDDistAcc;
            \addplot[geomstats, numpy] table [x index=0, y=geomstats_numpy] \ManSPDDistAcc;
            \addplot[geomstats, pytorch] table [x index=0, y=geomstats_pytorch] \ManSPDDistAcc;
            \addplot[geomstats, tensorflow] table [x index=0, y=geomstats_tensorflow] \ManSPDDistAcc;
            \addplot[geoopt] table [x index=0, y=geoopt] \ManSPDDistAcc;
            \addplot[manifolds] table [x index=0, y=manifolds] \ManSPDDistAcc;
            \addplot[manopt] table [x index=0, y=manopt] \ManSPDDistAcc;
            \addplot[pymanopt] table [x index=0, y=pymanopt] \ManSPDDistAcc;
            \addplot[riemopt] table [x index=0, y=riemopt] \ManSPDDistAcc;
            \nextgroupplot[title=exponential map]
            \addplot[maxacc] table [x=dim, y=maxacc] \ManSPDExpAcc;
            \addplot[geomstats, autograd] table [x index=0, y index=1] \ManSPDExpAcc;
            \addplot[geomstats, numpy] table [x index=0, y=geomstats_numpy] \ManSPDExpAcc;
            \addplot[geomstats, pytorch] table [x index=0, y=geomstats_pytorch] \ManSPDExpAcc;
            \addplot[geomstats, tensorflow] table [x index=0, y=geomstats_tensorflow] \ManSPDExpAcc;
            \addplot[geoopt] table [x index=0, y=geoopt] \ManSPDExpAcc;
            \addplot[manifolds] table [x index=0, y=manifolds] \ManSPDExpAcc;
            \addplot[manopt] table [x index=0, y=manopt] \ManSPDExpAcc;
            \addplot[pymanopt] table [x index=0, y=pymanopt] \ManSPDExpAcc;
            \addplot[riemopt] table [x index=0, y=riemopt] \ManSPDExpAcc;
            \addplot[roptlib] table [x index=0, y=roptlib] \ManSPDExpAcc;
            \nextgroupplot[title=logarithmic map]
            \addplot[maxacc] table [x=dim, y=maxacc] \ManSPDLogAcc;
            \addplot[geomstats, autograd] table [x index=0, y index=1] \ManSPDLogAcc;
            \addplot[geomstats, numpy] table [x index=0, y=geomstats_numpy] \ManSPDLogAcc;
            \addplot[geomstats, pytorch] table [x index=0, y=geomstats_pytorch] \ManSPDLogAcc;
            \addplot[geomstats, tensorflow] table [x index=0, y=geomstats_tensorflow] \ManSPDLogAcc;
            \addplot[geoopt] table [x index=0, y=geoopt] \ManSPDLogAcc;
            \addplot[manifolds] table [x index=0, y=manifolds] \ManSPDLogAcc;
            \addplot[manopt] table [x index=0, y=manopt] \ManSPDLogAcc;
            \addplot[pymanopt] table [x index=0, y=pymanopt] \ManSPDLogAcc;
            \addplot[riemopt] table [x index=0, y=riemopt] \ManSPDLogAcc;
        \end{groupplot}
        \matrix[
            matrix of nodes,
            anchor=south,
            draw=none,
            inner sep=0.2em,
        ] at (current bounding box.north) {
        {Accuracy on $\mathcal P(n)$} \\
        };
    \end{tikzpicture}
        \begin{tikzpicture}
        \begin{groupplot}[
            group style={
                group size=3 by 1,
                horizontal sep=3pt,
                y descriptions at=edge left,
            },
            axis x line*=bottom,
            axis y line*=left,
            xlabel=dimension $n$,
            ylabel={mean absolute error},
            log origin=infty,
            xmode=log,ymode=log,
            xtick = {2,3,4,8,16,32},
            xticklabels = {$2$, $3$, $4$, $8$, $16$,$32$},
            xmin=1.9, xmax=35,
            width=.41\textwidth,
            ylabel shift = -.7em,
            ymin=2e-17, ymax=1e-3,
        ]
            \nextgroupplot[title=distance]
            \addplot[maxacc] table [x=dim, y=maxacc] \ManPowDistAcc;
            \addplot[geomstats, autograd] table [x index=0, y index=1] \ManPowDistAcc;
            \addplot[geomstats, numpy] table [x index=0, y=geomstats_numpy] \ManPowDistAcc;
            \addplot[geomstats, pytorch] table [x index=0, y=geomstats_pytorch] \ManPowDistAcc;
            \addplot[geomstats, tensorflow] table [x index=0, y=geomstats_tensorflow] \ManPowDistAcc;
            \addplot[geoopt] table [x index=0, y=geoopt] \ManPowDistAcc;
            \addplot[manifolds] table [x index=0, y=manifolds] \ManPowDistAcc;
            \addplot[manopt] table [x index=0, y=manopt] \ManPowDistAcc;
            \addplot[pymanopt] table [x index=0, y=pymanopt] \ManPowDistAcc;
            \addplot[riemopt] table [x index=0, y=riemopt] \ManPowDistAcc;
            \nextgroupplot[title=exponential map]
            \addplot[maxacc] table [x=dim, y=maxacc] \ManPowExpAcc;
            \addplot[geomstats, autograd] table [x index=0, y index=1] \ManPowExpAcc;
            \addplot[geomstats, numpy] table [x index=0, y=geomstats_numpy] \ManPowExpAcc;
            \addplot[geomstats, pytorch] table [x index=0, y=geomstats_pytorch] \ManPowExpAcc;
            \addplot[geomstats, tensorflow] table [x index=0, y=geomstats_tensorflow] \ManPowExpAcc;
            \addplot[geoopt] table [x index=0, y=geoopt] \ManPowExpAcc;
            \addplot[manifolds] table [x index=0, y=manifolds] \ManPowExpAcc;
            \addplot[manopt] table [x index=0, y=manopt] \ManPowExpAcc;
            \addplot[pymanopt] table [x index=0, y=pymanopt] \ManPowExpAcc;
            \addplot[riemopt] table [x index=0, y=riemopt] \ManPowExpAcc;
            \addplot[roptlib] table [x index=0, y=roptlib] \ManPowExpAcc;
            \nextgroupplot[title=logarithmic map]
            \addplot[maxacc] table [x=dim, y=maxacc] \ManPowLogAcc;
            \addplot[geomstats, autograd] table [x index=0, y index=1] \ManPowLogAcc;
            \addplot[geomstats, numpy] table [x index=0, y=geomstats_numpy] \ManPowLogAcc;
            \addplot[geomstats, pytorch] table [x index=0, y=geomstats_pytorch] \ManPowLogAcc;
            \addplot[geomstats, tensorflow] table [x index=0, y=geomstats_tensorflow] \ManPowLogAcc;
            \addplot[geoopt] table [x index=0, y=geoopt] \ManPowLogAcc;
            \addplot[manifolds] table [x index=0, y=manifolds] \ManPowLogAcc;
            \addplot[manopt] table [x index=0, y=manopt] \ManPowLogAcc;
            \addplot[pymanopt] table [x index=0, y=pymanopt] \ManPowLogAcc;
            \addplot[riemopt] table [x index=0, y=riemopt] \ManPowLogAcc;
        \end{groupplot}
        \matrix[
            matrix of nodes,
            anchor=south,
            draw=none,
            inner sep=0.2em,
        ] at (current bounding box.north) {
        {Accuracy on $(\mathcal P(n))^{128\times 128}$} \\
        };
    \end{tikzpicture}
    \caption{Mean absolute error of distance (left), exponential maps (middle), and logarithmic maps (right) on the manifolds of rotations (top), manifold of symmetric positive definite matrices (middle), and its power manifold (bottom).
    The different colors correspond to the different software packages, where for Geomstats additionally the different line styles refer to different backends.}
    \label{fig:accuracy2}
\end{figure}

\section{Library-specific Details for the Speed and Accuracy Benchmark}\label{app:TechnicalDetails}

For \lstinline!Manifolds.jl!, version 0.8.57 of the library
and \lstinline!ManifoldsBase.jl! version 0.14.5 was used.
\begin{itemize}
    \item $\Sn{n}$: \lstinline!Sphere(n)!
    \item $\Rn{n}$: \lstinline!Euclidean(n)!
    \item $\Hn{n}$: \lstinline!Hyperbolic(n)!
    \item $\symposdef{n}$: \lstinline!SymmetricPositiveDefinite(n)!
    \item $\symposdef{n}^{128\times 128}$: \\ \lstinline!PowerManifold(SymmetricPositiveDefinite(n), ArrayPowerRepresentation(), 128, 128)!
    \item $\SOn{n}$: \lstinline!Rotations(n)!
\end{itemize}
Statically sized arrays from the \lstinline!StaticArrays.jl! library were used to represent points and vectors for $\Sn{n}$, $\Rn{n}$ and $\Hn{n}$ when $n \leq 16$, for  $\symposdef{n}$ when $n \in \{2, 3\}$ and for $\SOn{n}$ when $n \leq 4$.
For $\symposdef{n}^{128\times 128}$, the array type from \lstinline!HybridArrays.jl!\footnote{see \url{https://github.com/JuliaArrays/HybridArrays.jl}} version 0.4.15 was used with the first two axes statically sized.
All other cases used the standard \lstinline!Array! type.

All Python libraries were tested with Python version 3.9.16.final.0 (64 bit).

Geomstats version 2.5.0 was used with Autograd 1.5.0, NumPy 1.24.3, PyTorch 2.0.1, and Tensorflow 2.12.1.
The following commands were used to construct manifolds:
\begin{itemize}
    \item $\Sn{n}$: \lstinline[language=Python]!hypersphere.HypersphereMetric(n)!
    \item $\Rn{n}$: \lstinline[language=Python]!euclidean.EuclideanMetric(n)!
    \item $\Hn{n}$: \lstinline[language=Python]!hyperbolic.Hyperbolic(n).metric!
    \item $\symposdef{n}$: \lstinline[language=Python]!spd_matrices.SPDMetricAffine(n, 1)!
    \item $\symposdef{n}^{128\times 128}$:\\
    \lstinline[language=Python]!M = SPDMatrices(3)!\\
    \lstinline[language=Python]!M.metric.shape = (3, 3) # fixes an issue in geomstats!\\
    \lstinline[language=Python]!MPm = NFoldMetric(M.metric, 128*128)!
    \item $\SOn{n}$: \lstinline[language=Python]!special_orthogonal.SpecialOrthogonal(n)!
\end{itemize}
All manifolds were taken from the \lstinline[language=Python]!geomstats.geometry! path (omitted in the list above).

Geoopt version 0.5.0 was benchmarked.
The following commands were used to construct manifolds:
\begin{itemize}
    \item $\Sn{n}$: \lstinline[language=Python]!Sphere(torch.ones([n+1,n+1]))!
    \item $\Rn{n}$: \lstinline[language=Python]!Euclidean(1)!
    \item $\Hn{n}$: \lstinline[language=Python]!lorentz.math!
\end{itemize}
Manifolds listed above were taken from the \lstinline[language=Python]!geoopt.manifolds! path (omitted in the list above).
Other tested manifolds are not available in Geoopt.

Riemopt version 0.1.2 was benchmarked.
The following commands were used to construct manifolds:
\begin{itemize}
    \item $\Sn{n}$: \lstinline[language=Python]!Sphere()!
    \item $\Rn{n}$: \lstinline[language=Python]!Euclidean()!
    \item $\Hn{n}$: \lstinline[language=Python]!Hyperboloid()!
    \item $\symposdef{n}$: \lstinline[language=Python]!SPDAffineInvariant()!
    \item $\symposdef{n}^{128\times 128}$ \lstinline[language=Python]!SPDAffineInvariant()!
    \item $\SOn{n}$: \lstinline[language=Python]!SpecialOrthogonal()!
\end{itemize}
All manifolds were taken from the \lstinline[language=Python]!tensorflow_riemopt.manifolds! path (omitted in the list above).

Manopt version 7.1.0 (released on September 30, 2022) was used with Matlab version R2023a.
The following commands were used to construct manifolds:
\begin{itemize}
    \item $\Sn{n}$: \lstinline[language=Python]!spherefactory(n+1)!
    \item $\Rn{n}$: \lstinline[language=Python]!euclideanfactory(n)!
    \item $\Hn{n}$: \lstinline[language=Python]!hyperbolicfactory(n+1)!
    \item $\symposdef{n}$: \lstinline[language=Python]!sympositivedefinitefactory(n)!
    \item $\symposdef{n}^{128\times 128}$: \lstinline[language=Python]!powermanifold(sympositivedefinitefactory(n), 128*128)!
    \item $\SOn{n}$: \lstinline[language=Python]!rotationsfactory(n)!
\end{itemize}

Pymanopt version 2.1.1 was used.
The following commands were used to construct manifolds:
\begin{itemize}
    \item $\Sn{n}$: \lstinline[language=Python]!Sphere(n+1)!
    \item $\Rn{n}$: \lstinline[language=Python]!euclidean.Euclidean(n)!
    \item $\symposdef{n}$: \lstinline[language=Python]!PositiveDefinite(n)!
    \item $\symposdef{n}^{128\times 128}$: \lstinline[language=Python]!PositiveDefinite(n, 128 * 128), 128*128)!
    \item $\SOn{n}$: \lstinline[language=Python]!rotations.Rotations(n)!
\end{itemize}
Manifolds listed above were taken from the \lstinline[language=Python]!Pymanopt.manifolds! path (omitted above).

ROPTLIB version 0.8 (released on August 11, 2020) was used, compiled using GCC version 9.3.0 using the provided makefile.
The following commands were used to construct manifolds:
\begin{itemize}
    \item $\Sn{n}$: \lstinline[language=C++]!Sphere(n+1)!
    \item $\Rn{n}$: \lstinline[language=C++]!Euclidean(n, "double")!
    \item $\symposdef{n}$: \lstinline[language=C++]!SPDManifold(n)!
    \item $\symposdef{n}^{128\times 128}$: \lstinline[language=C++]!ProductManifold prodMani(1, &spdMani, 128 * 128);!, where the variable \lstinline[language=C++]!spdMani! was defined as \lstinline[language=C++]!SPDManifold spdMani(n);!
    \item $\SOn{n}$ is not directly implemented (the manual suggests using a special case of Stiefel instead) and was thus not tested
\end{itemize}
Hyperbolic manifold was not available in ROPTLIB. The tested version of ROPTLIB does not provide logarithmic maps, and distance calculation can only be performed as a norm of a tangent vector and were excluded as well.
\end{document}